\documentclass[journal,twoside,web]{ieeecolor}
\usepackage{tmi}
\usepackage{cite}
\usepackage{amsmath,amssymb,amsfonts}
\usepackage{algorithmic}
\usepackage{graphicx}
\DeclareGraphicsExtensions{.pdf,.jpeg,.png} 
\usepackage{textcomp}
\def\BibTeX{{\rm B\kern-.05em{\sc i\kern-.025em b}\kern-.08em
		T\kern-.1667em\lower.7ex\hbox{E}\kern-.125emX}}
\usepackage{array}
\usepackage{subcaption}
\usepackage{todonotes}
\usepackage{hyperref}

\usepackage{dsfont,pifont}

\begin{document}
	

\newcommand{\MetaModel}[1]		{\ensuremath{\textit{P}_\textit{#1}}}
\newcommand{\DensityModel}   	{\ensuremath{\textit{D}}} 
\newcommand{\FindingsModel}   	{\ensuremath{\textit{F}}} 
\newcommand{\LocalizationModel}	{\ensuremath{\textit{L}}} 
\newcommand{\SetModels}			{\ensuremath{\mathcal{M}}}

\newcommand{\SetViews}			{\ensuremath{\mathcal{I}_i}}

\newcommand{\score}[2]			{\ensuremath{p_{#1}^{#2}}}
\newcommand{\feature}[2]		{\ensuremath{\textit{feat}_{#1}^{#2}}}
\newcommand{\fusionvec}[1]		{\ensuremath{w_{#1}}}

\newcommand{\cmark}{\ding{51}}%

\title{Multi-task fusion for improving mammography screening data classification}

\author{Maria~Wimmer,
        Gert~Sluiter,
        David Major,
        Dimitrios Lenis,
		Astrid Berg,
        Theresa Neubauer,
        and~Katja~B\"{u}hler%
\thanks{The authors are with VRVis Zentrum f\"{u}r Virtual Reality und Visualisierung Forschungs-GmbH, 1220 Vienna, Austria (Corresponding author: Maria Wimmer, mwimmer@vrvis.at).}
\thanks{VRVis is funded by BMK, BMDW, Styria, SFG, Tyrol, and Vienna Business Agency in the scope of COMET - Competence Centers for Excellent Technologies (879730) which is managed by FFG. Thanks go to our project partner AGFA HealthCare for valuable input.}
}

\newcommand\copyrighttext{%
	\footnotesize © 2021 IEEE. Personal use of this material is permitted. Permission from IEEE must be obtained for all other uses, in any current or future media, including reprinting/republishing this material for advertising or promotional purposes, creating new collective works, for resale or redistribution to servers or lists, or reuse of any copyrighted component of this work in other works.}
\newcommand\mycopyrightnotice{%
	\begin{tikzpicture}[remember picture,overlay]
		\node[yshift=-15pt,xshift=-5pt] at (current page.north) {\fbox{\parbox{\dimexpr 20.5cm \relax }{\copyrighttext}}};
	\end{tikzpicture}%
}

\maketitle

\setlength{\fboxrule}{0pt}
\mycopyrightnotice

\begin{abstract} 
Machine learning and deep learning methods have become essential for computer-assisted prediction in medicine, with a growing number of applications also in the field of mammography. 
Typically these algorithms are trained for a \textit{specific task}, e.g., the classification of lesions or the prediction of a mammogram's pathology status. To obtain a comprehensive view of a patient, models which were all trained for the \textit{same task(s)} are subsequently ensembled or combined. 
In this work, we propose a pipeline approach, where we first train a set of \textit{individual, task-specific models} and subsequently investigate the fusion thereof, which is in contrast to the standard model ensembling strategy. We fuse model predictions and high-level features from deep learning models with \textit{hybrid patient models} to build stronger predictors on patient level. To this end, we propose a multi-branch deep learning model which efficiently fuses features across different tasks and mammograms to obtain a comprehensive patient-level prediction. 
We train and evaluate our full pipeline on public mammography data, i.e., DDSM and its curated version CBIS-DDSM, and report an AUC score of 0.962 for predicting the presence of any lesion and 0.791 for predicting the presence of malignant lesions on patient level. Overall, our fusion approaches improve AUC scores significantly by up to 0.04 compared to standard model ensembling. 
Moreover, by providing not only global patient-level predictions but also task-specific model results that are related to radiological features, our pipeline aims to closely support the reading workflow of radiologists.

\end{abstract}

\begin{IEEEkeywords}
Mammography, DDSM, CBIS-DDSM, Deep Learning, Model Fusion.
\end{IEEEkeywords}

\section{Introduction}

\IEEEPARstart{B}{reast} cancer is the most common cancer type in women and also the leading cause of death by cancer in women worldwide~\cite{cancer2020}. Fortunately, the mortality rate declined in recent years, one reason being the higher rate of early diagnosis due to the establishment of screening programs. Important cancer risk factors, such as breast density, can be detected and monitored early with such programs~\cite{cancer2020,destounis2020}.

Due to the increasing amount of imaging data, machine learning, especially deep learning algorithms are being developed to automatically process mammography data. Such models perform, for example, localization and classification of lesions~\cite{ribli2018,kooi2017}, breast density classification~\cite{lehman2019,kaiser2019}, or cancer risk prediction~\cite{mckinney2020short,yala2019}. These automated methods can be used for accelerating reading workflows \cite{kyono2020,ou2021}, or ideally, to support radiologists in their image interpretation and diagnosis~\cite{barnett2021}. 
Several recent studies report higher accuracies when combining AI algorithms with the assessment of a single radiologist~\cite{schaffter2020short} or improved performance of radiologists when aided by an AI system~\cite{kim2020short,rodriguezruiz2019}. 
Besides the obtained performance gains, the assistance of radiologists as well as \textit{human-computer collaboration} are becoming increasingly important aspects and challenges for future application in clinical practice~\cite{ou2021,geras2019}. To increase trust in AI support tools, not only the interpretability of black box models is being intensively studied~\cite{ribeiro2016,barredoarrieta2020short,shengeras2021short} but also the potential of providing intermediate model results that are linked to radiological features~\cite{barnett2021,kyono2020}. Recent user studies in cancer screening and diagnosis showed that clinicians profited more from models that provide detailed results compared to solutions delivering solely a benign/malignant assessment~\cite{tschandl2020short,cai2019}.

\begin{figure}
	\centering
	\includegraphics[width=1.0\linewidth]{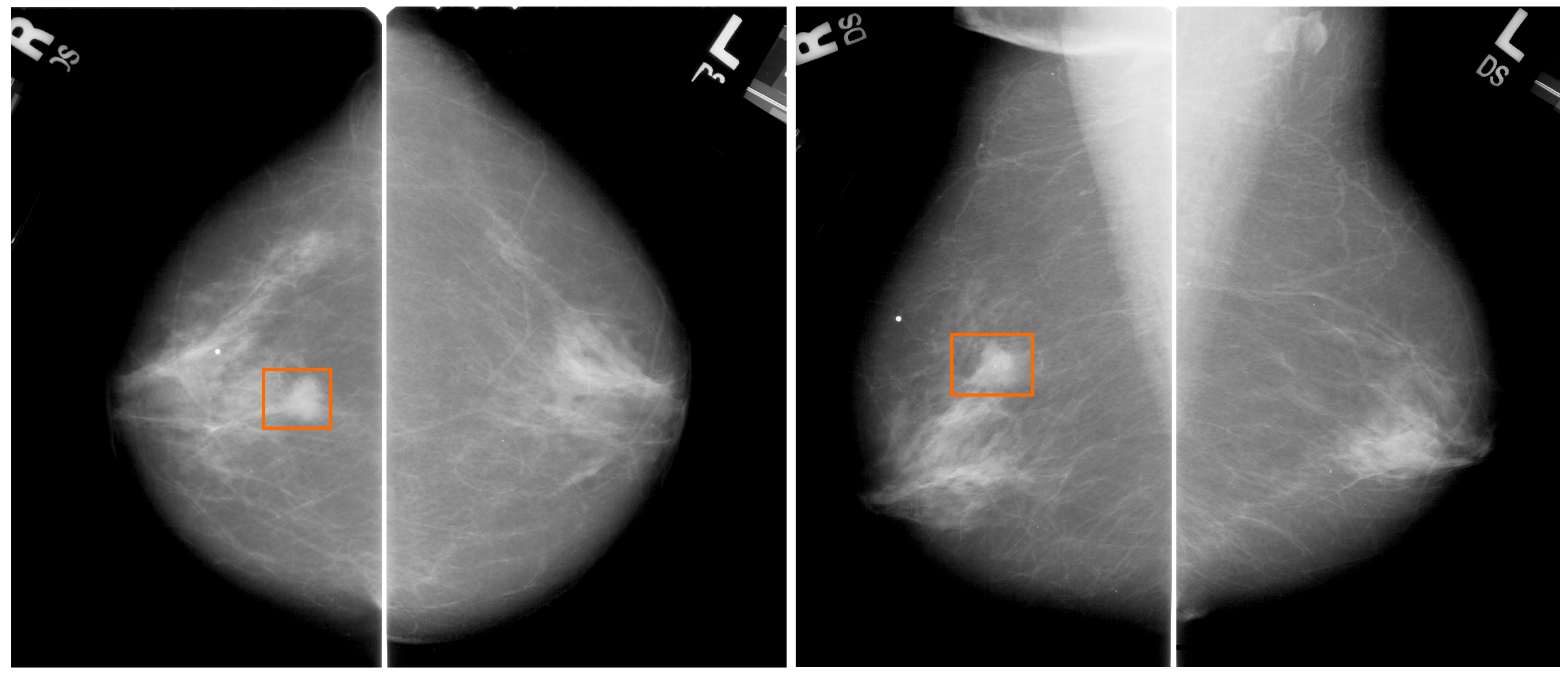}
	\caption{Standard mammography study of a patient showing the four standard views (from left to right): R-CC, L-CC, R-MLO, and L-MLO. The patient has a malignant mass in the right breast, highlighted in orange. Best viewed in color.}
	\label{fig:mammo_sample_case}
\end{figure}

\subsection{Related Work}
\label{sec:related work}

A standard mammography study (see Fig.~\ref{fig:mammo_sample_case}) comprises four X-ray images that correspond to two different imaging views from each breast: L-CC, R-CC, L-MLO, and R-MLO. Thereby, CC corresponds to the craniocaudal (CC) view, MLO to the mediolateral oblique (MLO) view, and L and R indicate the left or right breast, respectively. Radiologists analyze each view in detail and compare them to obtain a comprehensive view of a patient and render a diagnostic decision~\cite{sechopoulos2021}. Suspicious lesions, for example, can be visible in one view of a breast but may be obscured in the other view. Therefore, a thorough analysis is necessary. Various deep learning-based methods have been presented in the past years that analyze single- or multiple-view images at a time. However, this is strongly dependent on their task and related clinical question.

\subsubsection{Breast Density Scoring}
\label{sec:breast density related work}

Breast density is an important risk factor as dense breast tissue is related to the development of cancer. Furthermore, microcalcifications (MCs) and masses are harder to see on the mammograms, causing misdiagnoses~\cite{destounis2020}. The BI-RADS standard~\cite{sickles2013} defines density in four categories (a-d) as a measure of the breast tissue composition: ``almost entirely fatty``~(a), ``scattered areas of fibroglandular density``~(b), ``heterogeneously dense``~(c), and ``extremely dense``~(d). Assessment by radiologists usually has a high inter-observer variability~\cite{sprague2016short} due to the qualitative description of the four categories. Therefore, automated density classification models often focus on the two superclasses \textit{not dense} or \textit{fatty}~(a+b) and \textit{dense} (c+d)~\cite{kaiser2019}. 

Recent works utilized all four mammography views via multi-view CNNs to classify breast density into the four density categories~\cite{wu2018short} or in both superclasses~\cite{wu2018short,kaiser2019}. In contrast, Lehman et al.~\cite{lehman2019} trained a ResNet-18 model to classify single mammograms and assigned the consensus density across all views for the patient. Other methods used a refined AlexNet to classify the two middle density classes (b+c)~\cite{mohamed2018} or performed unsupervised feature learning to segment dense tissue and derive a density scoring per image~\cite{kallenberg2016short}.

\subsubsection{Lesion Localization and Classification}
\label{sec:lesion deteciton related work}

\begin{figure}
	\centering
	\includegraphics[width=\linewidth]{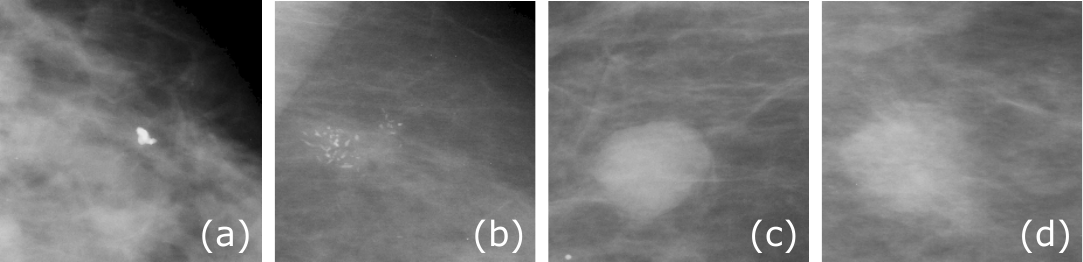}
	\caption{Patches showing a benign calcification (a), malignant MC cluster (b), benign mass (c), and a malignant mass (d).}
	\label{fig:sample_patches_lesion}
\end{figure}

Exact localization and classification of lesions (i.e., masses, calcifications, and clusters of MCs) in mammograms are crucial as they are important risk factors or already indicators of cancer~\cite{sechopoulos2021} (see Fig.~\ref{fig:sample_patches_lesion}). While many works perform lesion localization, quantification, classification, or all together~\cite{ribli2018,kooi2017,kooi2017_differences,carneiro2017,akselrodballin2017}, others solely classify already extracted lesions on patches~\cite{arevalo2016,samala2017,mordang2016,dgani2018,barnett2021}.
The use of classical feature extraction and machine learning methods, or the combination thereof with CNNs, has been intensively investigated in the literature~\cite{elnaqa2002,arevalo2016,kooi2017,kooi2017_differences,agarwal2019}.
Mordang et al.~\cite{mordang2016} were the first to use CNNs for MC localization and utilized a VGG-like architecture for this task. Various studies focused on the classification of MCs and MC clusters~\cite{wang2018,shachor2020,dgani2018}, e.g., with a combination of a difference-of-Gaussians detector and two-stream CNNs~\cite{wang2018}. 
Dhungel et al.~\cite{dhungel2017}, among many others~\cite{arevalo2016,samala2017,agarwal2019,almasni2018short,barnett2021}, performed localization and analysis of masses. They combined deep belief networks, Gaussian mixture models, and CNNs for mass detection. Barnett et al.~\cite{barnett2021} recently proposed an interpretable mass classification framework with the goal to follow the reasoning process of radiologists.
Finally, state-of-the-art object detection approaches like Faster R-CNN~\cite{ren2015} or YOLO have been applied for lesion localization and classification~\cite{ribli2018,agarwal2020,akselrodballin2017,almasni2018short,lotter2021short}. Ribli et al.~\cite{ribli2018} utilized a Faster R-CNN with a VGG16 backbone to detect and classify lesions into malignant and benign classes individually. %
Others extended a Faster R-CNN model by a cascaded classification step to reduce false-positive detected lesions~\cite{akselrodballin2017}.

\subsubsection{Malignancy Scoring}

Several studies classify single or multiple mammograms directly to obtain a score assessing whether the view image is cancerous~\cite{shen2019,lotter2017,mckinney2020short,shengeras2021short,lotter2021short} or contains a (specific) malignant or benign finding~\cite{zhu2017,wu2019short,shu2020,tardymateus2021}. Recent works utilized, e.g., an all-convolutional design combined with curriculum learning~\cite{shen2019}, multi-instance learning~\cite{zhu2017,lotter2021short,shengeras2019,shengeras2021short}, self-supervised methods~\cite{tardymateus2021}, or a multi-view-multi-task approach~\cite{kyono2020}. Wu et al.~\cite{wu2019short} concatenated heatmaps obtained from sliding window patch classification to classify full images. 
Other works derive a malignancy score per view image, breast, or patient by averaging or considering the maximum score, e.g., obtained from a Faster R-CNN~\cite{ribli2018} or a map of pixelwise abnormality scores~\cite{kim2020short}.

\subsubsection{Feature or Information Fusion}
\label{sec:feature fusion related work}

The fusion of \textit{features} or, more generally, of (extracted) \textit{information} is inspired by how radiologists assess and compare ROIs and mammograms to obtain a comprehensive view of a patient. The term \textit{feature} can refer to ``classical, handcrafted`` features (e.g., gabor filters, curvelets, entropy, etc.), CNN-features extracted by a CNN, or non-imaging features (e.g., patient age). 
The extraction and fusion can be performed at different scales, for example, \textit{locally} from/within a single-view image, patches or across ROIs~\cite{shengeras2019,shengeras2021short,lotter2017,zhu2017,lotter2021short,kooi2017,dgani2018}. Kooi et al.~\cite{kooi2017} fused CNN and classical features extracted from patches within a single mammogram. 
Lotter et al.~\cite{lotter2017} and Shen et al.~\cite{shengeras2021short} fused local CNN patch features, whereas the former extracted them with a sliding window approach, and the latter extracted only CNN features from salient regions obtained with a global image classifier. 
Another common approach is to utilize \textit{multiple views} for localization and classification of lesions and full images, as summarized by Jouirou et al.~\cite{jouirou2019}. While Shachor et al.~\cite{shachor2020} dynamically combined classical features from local patches from MLO and CC view for calcification classification, Kooi et al.~\cite{kooi2017_differences} fused CNN features from ROIs across views for malignant mass detection. The usage of \textit{multi-view CNNs}, where each view image is processed with a CNN, followed by feature fusion at a given layer, has been studied as well~\cite{carneiro2017,geras2018short,kyono2019} for different purposes, e.g., BI-RADS scoring~\cite{geras2018short} or breast density classification~\cite{kaiser2019,wu2018short}. McKinney et al.~\cite{mckinney2020short} developed several models, which used different fusion and combination strategies, e.g., concatenation of CNN patch features across all views, fusion of CNN image-level features per breast, and/or patient, or concatenation of non-imaging features (e.g., patient age) with CNN features. 

The last stage is \textit{decision-level fusion}, i.e., fusion of predictions, which has been investigated by Kyono et al.~\cite{kyono2019}, for example. 
They predicted several radiological features (e.g., breast density, diagnosis, age) with a multi-task CNN separately for each view, fused them, and classified the patient as benign or malignant.
Finally, the naive ensembling of predictions from different models, e.g., via averaging, can also be interpreted as decision fusion~\cite{mckinney2020short,ribli2018,kim2020short}. 


\subsubsection{Summary}
\label{sec:related work summary}

While many recent works directly classify ROIs or view images with, e.g., CNNs, a significant part utilizes some form of information fusion when processing mammography data (see Table~\ref{tab:related_work_comparison}). The reasons are manifold: 
fusion is performed to \textit{(i)} incorporate different aspects at different levels (ROI, image, patient), \textit{(ii)} thus, increase robustness and performance of classification models~\cite{kooi2017,kooi2017_differences,shengeras2019,shengeras2021short,lotter2017,zhu2017,shachor2020,dgani2018,lotter2021short}, and \textit{(iii)} increase explainability and interpretability of model predictions~\cite{barnett2021,shengeras2021short,barredoarrieta2020short,kyono2019,kyono2020}. 
Methods that fuse predictions across one or more ROIs or mammograms usually build upon models that predict the same scores for the same task or perform standard model ensembling strategies~\cite{mckinney2020short,kyono2019,kyono2020,lehman2019}. 
On the other hand, methods that perform a fusion of features within or across images mostly do not provide intermediate results (e.g., assessment of suspicious regions) but only final classification results. 
Although recent user studies highlight the potential of providing detailed classification results or pinpointing to suspicious regions~\cite{tschandl2020short,cai2019}, only a few proof-of-concept studies explored fusion and the potential of providing intermediate results similar to the assessment of radiologists in the field of mammography~\cite{barnett2021,kyono2019,kyono2020}. These methods operated only on lesion-level~\cite{barnett2021} or fused models that predict the same multi-task scores~\cite{kyono2019,kyono2020}. 
To the best of our knowledge, the fusion of models trained for \textit{different} tasks is not being studied in the context of mammography.

\begin{table*}[t!]
	\centering
	\caption{Overview on related works. Target: density = breast density classification, lesions = lesion localization and/or classification, malignancy = prediction of BI-RADS, benign/malignant, cancer yes/no, etc., on image/patient level; Data: name of image database; Fusion: \cmark = some form of fusion involved; Intermediate / Sub-results: type of intermediate/additional results provided apart from final scores; Method: brief summary (loc. = localization, seg. = segmentation, class. = classification, RF~=~random forest, DBN = deep belief network, GMM = gaussian mixture model, DoG = difference of gaussian).}
	\label{tab:related_work_comparison}
	\begin{tabular}{|l|c|c|c|l|l|}
		\hline
		Author 							& Target 		& Data		& Fusion	& Intermediate / Sub-results	& Method \\
		\hline \hline
		\cite{wu2018short,kaiser2019}	& density		& private	& \cmark	& no							& multi-view CNN \\
		\hline
		\cite{lehman2019,mohamed2018}	& density		& private	& 			& no							& single-view CNN \\
		\hline
		\cite{kallenberg2016short} 		& density		& private	& \cmark	& dense tissue segmentation	& \parbox{5.3cm}{multi-scale unsupervised seg. + texture scoring} \\	
		\hline
		\cite{dhungel2017}	& lesions		& INbreast		& 			& no		& DBN + GMM (loc.), CNN + RF (class.)\\
		\hline
		\cite{elnaqa2002}$^1$, \cite{arevalo2016}$^2$	& lesions		& BCDR$^2$, private$^1$	& 			& no		& SVM$^1$ / CNN + SVM$^2$ for class. \\
		\hline
		\parbox{1.5cm}{\cite{samala2017}$^1$, \cite{mordang2016}$^2$,\\ \cite{agarwal2019}$^3$}		& lesions		& \parbox{3cm}{\centering DDSM$^1$, private$^{1,2}$, \\ CBIS-DDSM$^3$, INbreast$^3$}		& 			& no		& CNN for class.$^{1,3}$ / loc. + class.$^2$ \\
		\hline 
		\cite{wang2018}					& lesions		& private	& \cmark	& no		& DoG + multi-scale two-stream CNN \\
		\hline
		\cite{shachor2020}				& lesions		& DDSM		& \cmark	& no		& multi-view CNN \\
		\hline
		\cite{dgani2018}				& lesions		& DDSM		& \cmark	& no		& classical features + feed forward network \\
		\hline
		\cite{kooi2017}					& lesions		& private	& \cmark	& no 		& \parbox{5cm}{candidate loc. (RF) + class. (CNN features + classical texture features)} \\
		\hline
		\cite{kooi2017_differences} 	& lesions		& private	& \cmark	& no		& \parbox{5cm}{dual-stream CNN for lesion ROI class.}\\
		\hline
		\cite{barnett2021} 				& lesions		& private	& \cmark & \parbox{3.5cm}{class activation map,\\ mass margin class score} & \parbox{5cm}{case-based reasoning, compares parts of new images to learned prototypes} \\
		\hline
		\parbox{1.5cm}{\cite{ribli2018}$^\text{1}$, \cite{agarwal2020}$^\text{2}$\\ \cite{akselrodballin2017}$^\text{3}$, \cite{almasni2018short}$^\text{4}$}		& lesions	& \parbox{3cm}{\centering DDSM$^\text{1,4}$, INbreast$^\text{1,2,3}$,\\ OPTIMAM$^\text{2}$, private$^\text{1,3}$}	& 		& no		& \parbox{5cm}{Faster R-CNN/YOLO-based lesion\\localization + classification}  \\
		\hline
		\cite{lotter2021short}	& \parbox{1.5cm}{\centering lesions, \\malignancy}	& \parbox{3cm}{\centering DDSM, OPTIMAM,\\ private}	& \cmark	& \parbox{3.5cm}{benign + malignant lesions (bounding boxes)} & \parbox{5cm}{RetinaNet-based approach + multi-stage training (fully + weakly supervised, multi-instance learning)} \\
		\hline
		\cite{tardymateus2021}		& \parbox{1.5cm}{\centering lesions, \\malignancy}		& INbreast, private		& \cmark	& malignancy probability map & \parbox{5cm}{self- and weakly supervised reconstruction for lesion loc./seg., image-level class.} \\
		\hline
		\cite{carneiro2017}			& malignancy	& INbreast, DDSM 			& \cmark	& no	& multi-view CNN \\
		\hline
		\cite{shen2019} 			& malignancy	& CBIS-DDSM, INbreast 		& 		& salient regions & all-convolutional CNN (two-stage) \\
		\hline
		\cite{zhu2017} 				& malignancy 	& INbreast 		& \cmark 	& no 	& multi-instance approach \\
		\hline
		\cite{geras2018short,wu2019short}  	& malignancy	& private 		& \cmark 	& \parbox{3cm}{heatmaps of malignant /\\ benign+malignant regions}	& \parbox{5cm}{multi-view CNN(s), \\fusion at different stages} \\
		\hline
		\cite{shu2020}				& malignancy	& INbreast, CBIS-DDSM	& 	& malignant regions & \parbox{5.3cm}{CNN + region-based/global group-max\\ pooling}  \\
		\hline
		\cite{kim2020short} 		& malignancy	& OPTIMAM, private & 		&  pixel-wise abnormality score &	semi-supervised CNN (two-stage) \\
		\hline
		\cite{lotter2017}			& malignancy 	& DDSM		& \cmark	& no				& multi-scale CNN + curriculum learning\\
		\hline
		\cite{shengeras2021short}	& malignancy	& private	& \cmark	& \parbox{3.3cm}{saliency maps\\ (malignant findings)}	& \parbox{5cm}{weakly supervised approach,\\ global (weak loc.) + local CNN} \\
		\hline
		\cite{kyono2019,kyono2020}			& malignancy	& private	& \cmark	& \parbox{3.5cm}{radiological features per view, \\heatmaps}	& multi-view, multi-task CNN \\
		\hline
		\cite{mckinney2020short}	& cancer risk 	& \parbox{3cm}{\centering OPTIMAM, \\CBIS-DDSM, private} & \cmark & \parbox{3.5cm}{malignant regions\\ (bounding boxes)} & \parbox{5cm}{patch-level, image-level and CNN+non-imaging feature fusion (various models)}\\
		\hline
		This work					& \parbox{1.5cm}{\centering density, \\lesions, \\malignancy} & DDSM, CBIS-DDSM & \cmark &\parbox{3.5cm}{breast density, lesions (bounding box + label), findings classification} & \parbox{5cm}{task-specific CNNs (multiple scales), \\feature and prediction fusion with\\ CNNs + MLPs}\\
		\hline
	\end{tabular}
\end{table*}

\subsection{Contribution}
\label{sec:contribution}

Closing this gap, we investigate information fusion for mammography from another perspective by focusing on the fusion of features and predictions from \textit{individual, task-specific models} to obtain a comprehensive assessment on patient level. To this end, we propose a \textit{pipeline approach} comprising
\begin{itemize}
	\item the development of three \textit{task-specific models}, namely \textit{(i)} a breast density classification model, \textit{(ii)} a lesion localization model, \textit{(iii)} and a findings classifier, as a basis for fusion, and
	\item the investigation of two fusion strategies: \textit{(i)} the fusion of high-dimensional, task-specific CNN features with a \textit{multi-input embedding CNN} and \textit{(ii)} prediction score fusion of model predictions with MLPs.
\end{itemize}

By building upon task-specific features and decisions, we obtain \textit{hybrid patient meta-models}, which access these intermediate results in their prediction. Due to the two-stage nature of our method, we report not only a global score on patient level but make the sub-results that reflect radiological features also accessible to the clinician.

We train both fusion approaches for two different classification targets, which we will refer to as \textit{patient predictions} (i.e., the prediction of the respective model). We predict \textit{(i)} the presence of \textit{any} lesion (\textit{lesion prediction}), \textit{(ii)} and whether the patient has any \textit{malignant} lesion (\textit{malignancy prediction}).

At each stage in our pipeline, we aim for resource-efficient models, and therefore, utilize lightweight architectures like MobileNets~\cite{howard2017} for image classification-related tasks. 
The full pipeline was trained and evaluated on the well-known and publicly available DDSM~\cite{ddsm1998,ddsm2001} and CBIS-DDSM datasets~\cite{lee2016,lee2017}. In a comprehensive technical analysis, we show that our task fusion strategy improves patient-level classification over standard model ensembling. A detailed analysis of results and discussion thereof as well as future clinical perspectives are provided in the discussion.


\section{Materials and Methods}
\label{sec:methods}

We define a set of mammography images $\mathcal{I}_i = \{I^v_i\}$ for patient $i$ and mammography image view $v \in$ \{L-CC, L-MLO, R-CC, R-MLO\}. We will refer to this set $\SetViews$ as \textit{exam} or \textit{case} of patient $i$.

\subsection{Data}
\label{sec:data}

We utilize two publicly available mammography databases for our experiments: the Digital Database for Screening Mammography (DDSM) \cite{ddsm1998,ddsm2001} and its curated version CBIS-DDSM \cite{lee2016,lee2017}. 

\subsubsection{DDSM and CBIS-DDSM Dataset}
\label{sec:ddsm and cbis-ddsm}

The \textit{original DDSM dataset}~\cite{ddsm1998,ddsm2001} comprises 2620 mammography screening exams $\SetViews$, collected from four different sites acquired with four different scanners. The data is grouped in four categories:
\begin{itemize}
	\item \textit{normal} (695 cases): normal exams with no suspicious abnormalities and proven normal exams four years later
	\item \textit{benign without callback} (141 cases): cases with benign abnormality but without need for callback
	\item \textit{benign} (870 cases): including suspicious findings which were identified as benign findings after callback
	\item \textit{cancer} (914 cases): cancer was proven via histology
\end{itemize}

An expert radiologist labeled the breast density per patient and provided pixel-level annotation for abnormalities. Each abnormality is described following the BI-RADS standard~\cite{sickles2013}, including lesion type (mass or calcification) and further details like shape, lesion margin, and calcification type.

The \textit{CBIS-DDSM dataset}~\cite{lee2016,lee2017} was published at The Cancer Imaging Archive~\cite{tcia2013short} as curated version of the original DDSM set, whereby only images showing one or more lesions have been transferred. Annotated masses were re-checked by a radiologist, and pixel-wise annotations have been refined with an automated segmentation algorithm. However, annotations of calcifications remained unchanged. The authors also provided a predefined split into train and test sets to ensure comparability between methods evaluated on this dataset. Overall, the CBIS-DDSM dataset comprises 3568 annotated lesions (1696 masses, 1872 calcifications) in a total of 3032 mammography view images. For further details on the data, we refer to the original publications~\cite{lee2016,lee2017}.

\subsubsection{Data Harmonization and Preparation}
\label{sec:data harmonization}

While providing enhanced annotation quality, the CBIS-DDSM dataset has two shortcomings: first, the absence of normal images without lesions, and second, the lack of full patient mammography exams including all four views. To utilize both resources without losing their individual benefits, we prepare the data as follows:

First, we preprocess the DDSM set in the same way as it was done for the CBIS-DDSM data, including optical density normalization and remapping the data to the full 16-bit range~\footnote{\url{https://github.com/fjeg/ddsm_tools}}.

Next, we match, i.e., compare the CBIS-DDSM images to the preprocessed DDSM data to identify corresponding cases and obtain a total of 2590 full mammography exams. We assign the malignancy status of a lesion according to the curated annotation from CBIS-DDSM, whereby ``benign without callback`` will be treated as a benign case. 

Finally, we identify potential ambiguous cases which have been originally in the cancer, benign, or benign without callback subset in DDSM but have not been transferred to CBIS-DDSM. Since the status of the lesions for these 329 cases remains unclear, we exclude them. Further, we exclude seven additional exams, which are either incomplete, i.e., not all four views are present, or appeared with different imaging data and annotations in different subsets of DDSM and CBIS-DDSM. This leads to our final set comprising \textit{2254 cases}.

\subsubsection{Train, Validation, Test Split}
\label{sec:data split}

We split the dataset into train, validation, and test data on case-level and, thus, ensure that images from one case are not distributed across different sets. We preserve the train/test split of the data provided with the CBIS-DDSM set. The remaining normal cases are randomly distributed in the same ratio ($\sim$80\% training images) to the train/test set in a way that the distribution of breast density is similar in the three sets. From the obtained train set, we randomly select $\sim$12\% of cases for the validation set in a way that the ratio of different breast density classes, lesion types, and pathology is similar across the three sets (see Table~\ref{tab:data_distributions}). Overall, the train, validation, and test set comprise 1511, 290, and 453 cases, respectively. Out of the 2254 cases, 174 contain more than one lesion, with the maximum number of lesions per case being 24.

\begin{table}[]
	\centering
	\caption{Distribution of breast density, lesion type, and pathology status in train, validation, and test set.}
	\label{tab:data_distributions}
	\begin{tabular}{|ll|c|c|c|c|}
		\hline
		&                                                                       & Train                                                         & Validation                                                 & Test                                                        & Total                                                         \\ \hline
		\multicolumn{1}{|l|}{Density}   & \begin{tabular}[c]{@{}l@{}}a\\ b\\ c\\ d\end{tabular}                 & \begin{tabular}[c]{@{}c@{}}207\\ 567\\ 448\\ 289\end{tabular} & \begin{tabular}[c]{@{}c@{}}40\\ 108\\ 86\\ 56\end{tabular} & \begin{tabular}[c]{@{}c@{}}50\\ 176\\ 134\\ 93\end{tabular} & \begin{tabular}[c]{@{}c@{}}297\\ 851\\ 668\\ 438\end{tabular} \\ \hline
		\multicolumn{1}{|l|}{Lesion}    & \begin{tabular}[c]{@{}l@{}}normal\\ mass\\ calcification\end{tabular} & \begin{tabular}[c]{@{}c@{}}481\\ 583\\ 485\end{tabular}       & \begin{tabular}[c]{@{}c@{}}107\\ 93\\ 96\end{tabular}      & \begin{tabular}[c]{@{}c@{}}105\\ 201\\ 150\end{tabular}     & \begin{tabular}[c]{@{}c@{}}693\\ 877\\ 731\end{tabular}       \\ \hline
		\multicolumn{1}{|l|}{Pathology} & \begin{tabular}[c]{@{}l@{}}normal\\ benign\\ malignant\end{tabular}   & \begin{tabular}[c]{@{}c@{}}481\\ 522\\ 508\end{tabular}       & \begin{tabular}[c]{@{}c@{}}107\\ 98\\ 85\end{tabular}      & \begin{tabular}[c]{@{}c@{}}105\\ 199\\ 149\end{tabular}     & \begin{tabular}[c]{@{}c@{}}693\\ 819\\ 742\end{tabular}       \\ \hline
	\end{tabular}
\end{table}


\subsection{Task-specific Mammography Models}
\label{sec:task models}

The first stage in our pipeline is the development of a set $\SetModels$ of three resource-efficient, task-specific models $\SetModels~=~\{\DensityModel, \LocalizationModel, \FindingsModel\}$, which are the base for our patient model $\MetaModel{}$:
\begin{itemize}
	\item $\DensityModel$ performs breast density classification,
	\item $\LocalizationModel$ delivers bounding boxes around localized lesions and their respective class label, and
	\item $\FindingsModel$ predicts the presence/absence of lesions in an image.
\end{itemize}

\subsubsection{Breast Density Model ($\DensityModel$)}
\label{sec:density block}

Radiologists include all four view images $\SetViews$ in the assessment of a patient's breast density. Recent deep learning based density classification models follow this standard and utilize all views as input~\cite{wu2018short,kaiser2019}, whereas the usage of only one view has also been studied \cite{lehman2019}. We propose a two-stage approach where we employ both ideas in the design of density model $\DensityModel$ to increase robustness and classification performance.

\begin{figure}
	\centering
	\includegraphics[width=0.85\linewidth]{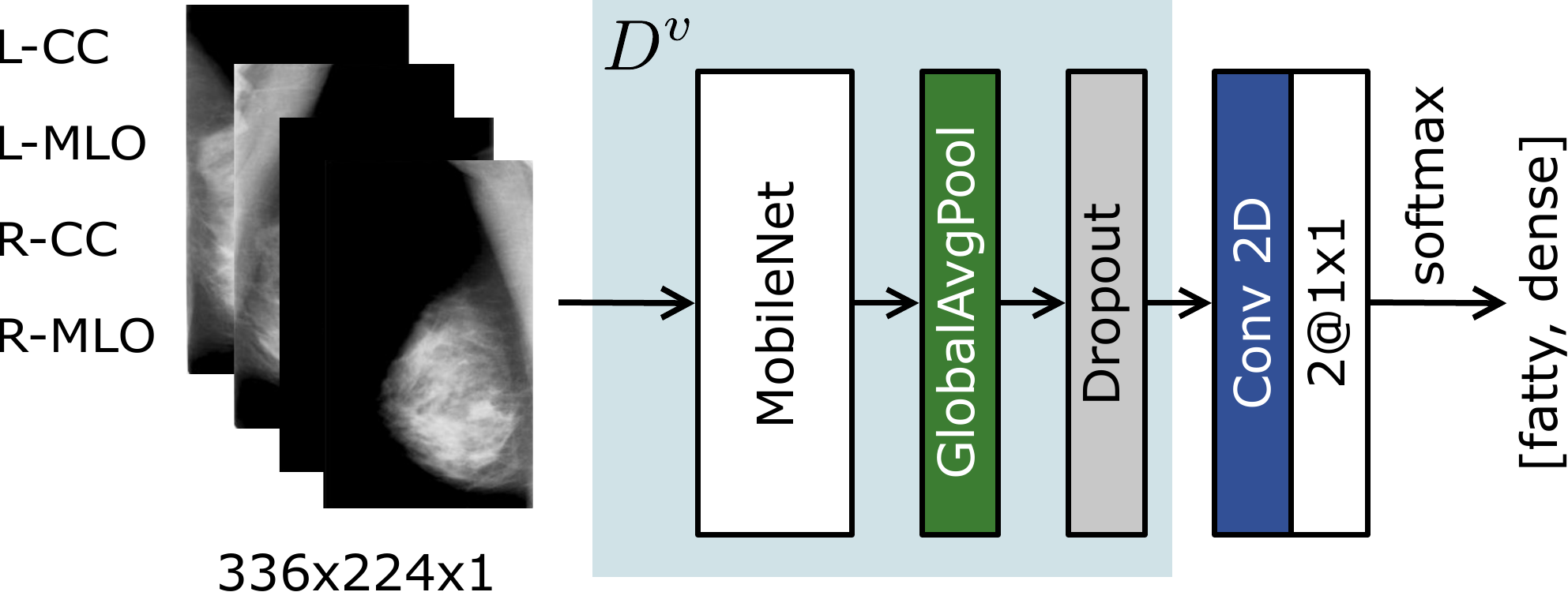}
	\caption{Density view model $\DensityModel^v$ for view $v \in$ \{L-CC, L-MLO, R-CC, R-MLO\}}
	\label{fig:density_view_model}
\end{figure}

\begin{figure}
	\centering
	\includegraphics[width=0.95\linewidth]{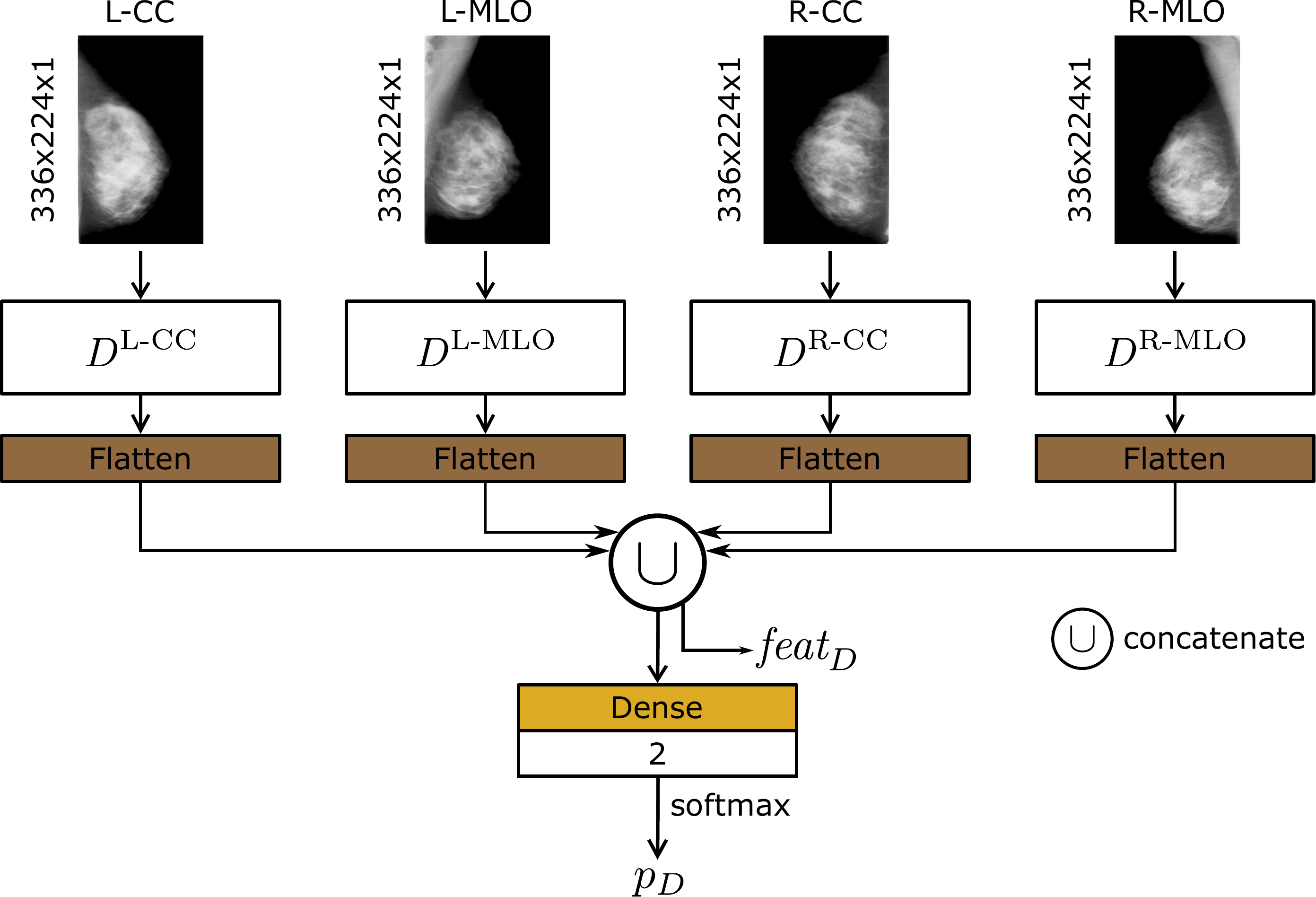}
	\caption{Density patient model $\DensityModel$}
	\label{fig:density_patient_model}
\end{figure}

We build a view model $\DensityModel^v$ first, which uses any single mammography image $I^v_i$ as input to predict the density superclass, i.e., \textit{fatty} or \textit{dense}. The model is built upon a MobileNet classifier~\cite{howard2017} with global average pooling, followed by a 1x1 convolution layer (see Fig.~\ref{fig:density_view_model}).
Our final model $\DensityModel$ takes the four standard mammography views $\SetViews$ as input where each image is passed to a separate branch (see Fig.~\ref{fig:density_patient_model}). Each view branch consists of a density view model $\DensityModel^v$, whereby the dropout rate is increased from 0.001 in model $\DensityModel^v$ to 0.5 in $\DensityModel$. After the following flattening operation, the 1D feature vectors are concatenated, and a final dense layer predicts the density superclass. The obtained density score $\score{\DensityModel}{}$ at patient-level depicts the score corresponding to the ``dense`` class.

\begin{figure}[!h]
	\centering
	\includegraphics[width=\linewidth]{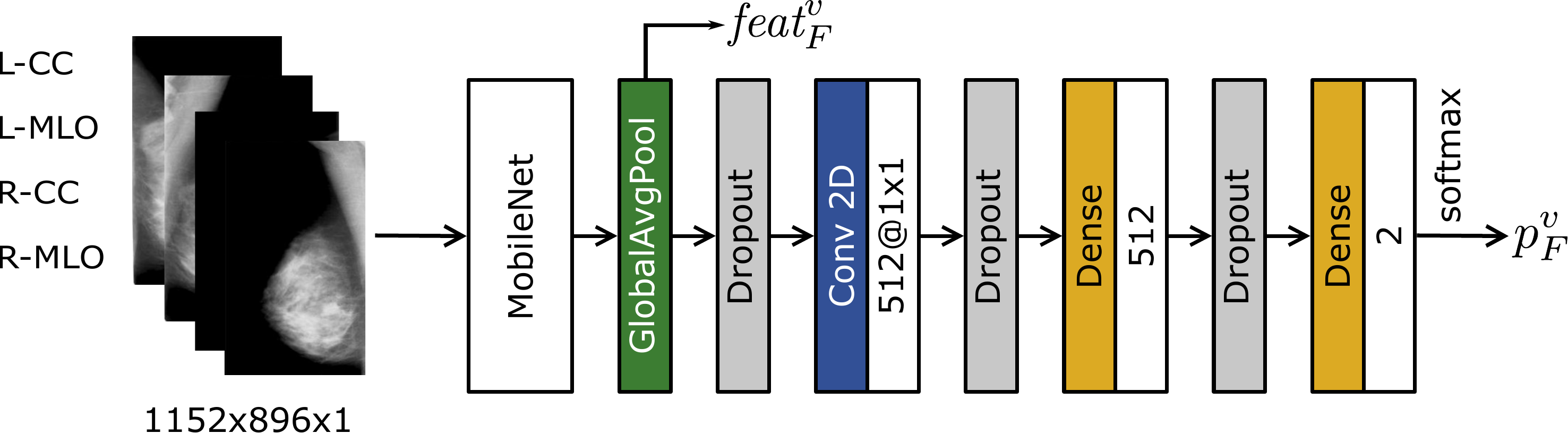}
	\caption{Findings model $\FindingsModel$}
	\label{fig:findings_model}
\end{figure}

\subsubsection{Findings Model ($\FindingsModel$)}
\label{sec:findings block}

The objective of this model is to classify any single-view image $I^v_i$ into ``normal`` or ``image containing \textit{any} findings``, i.e., lesions. Such a model could be, for example, integrated into a reporting system, in which images with lesions are examined first by a medical expert. Again, we aim for a resource-efficient model to solve this task, and thus, we extend on our previous work~\cite{major2020,lenis2020}, where we already successfully applied MobileNet~\cite{howard2017} in this context. Fig.~\ref{fig:findings_model} illustrates our findings model $\FindingsModel$ with a MobileNet feature extractor and a modified classifier on top. Adding an additional dense and dropout layer increased the classification accuracy and the generalization capability of the model. Additionally, we use an increased dropout rate of 0.5 to stronger regularize the network. 
The output for each view image $I^v_i$ is the score $\score{\FindingsModel}{v}$ which determines whether there is any lesion in $I^v_i$. 
 
\subsubsection{Localization Model ($\LocalizationModel$)}
\label{sec:detections block}

\begin{figure}
	\centering
	\includegraphics[width=0.95\linewidth]{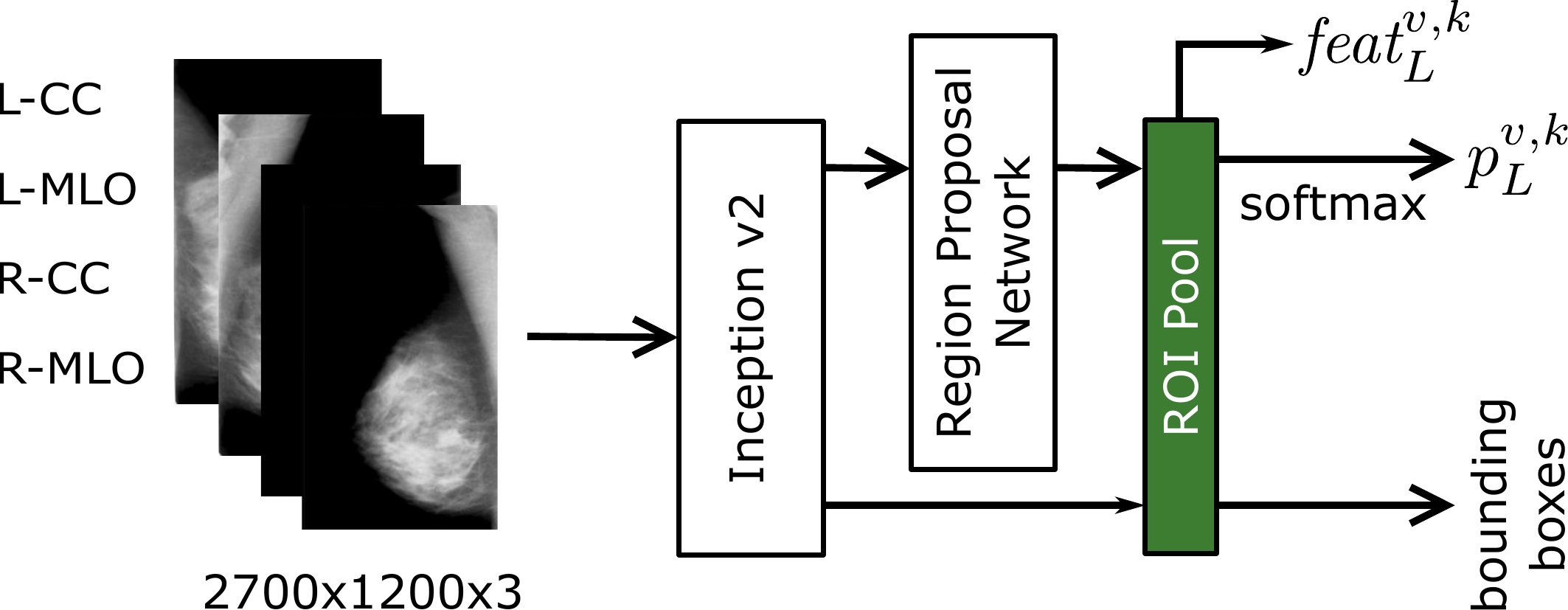}
	\caption{Localization model $\LocalizationModel$}
	\label{fig:detection_model}
\end{figure}

Similar to radiologists, we aim to detect the exact location of lesions within an image $I^v_i$ and classify them into their correct type and malignancy status. The localization and characterization of lesions are important tasks, as they can be risk factors or already indicators of cancer~\cite{sechopoulos2021}. Therefore, we develop model $\LocalizationModel$ to localize lesions and classify them in either ``benign calcification``, ``malignant calcification``, ``benign mass``, or ``malignant mass``.
Inspired by recent works on lesion localization~\cite{ribli2018,agarwal2020,akselrodballin2017}, we utilize the well-known Faster R-CNN~\cite{ren2015} architecture. InceptionV2~\cite{szegedy2016} serves as feature extractor, which was already successfully applied in the context of mammography lesion localization~\cite{agarwal2020}. 
Fig.~\ref{fig:detection_model} illustrates the architecture. Our localization model $\LocalizationModel$ classifies localized lesions into four types (benign calcification, malignant calcification, benign mass, and malignant mass) and assigns $k \in [0,n]$ scores $\score{\LocalizationModel}{v,k}$, depending on the number of detected lesions that are found in $I^v_i$.

\subsection{Patient Meta-Model ($\MetaModel{}$)}
\label{sec:meta model}

The hybrid patient meta-model $\MetaModel{}$ aims to efficiently combine the task-specific building blocks $\SetModels$ to obtain a comprehensive patient-level assessment while preserving the individual model predictions related to radiological features and risk factors. We consider two different patient predictions:
\begin{itemize}
	\item \textit{lesion prediction:} whether the patient has any lesion, regardless of pathology,
	\item \textit{malignancy prediction:} whether the patient is malignant, i.e., has any malignant lesion.
\end{itemize}

The fusion of different models can be performed at various stages, whereby, again, our goal is to develop \textit{resource-efficient} variants. For this, we compare the fusion of \textit{prediction scores} as well as the fusion of \textit{features} from the individual models.

\subsubsection{Fusion of Predictions ($\MetaModel{score}$)}

The three task-models $\SetModels$ deliver different prediction scores $\score{m}{} \in [0,1], m \in \SetModels$ at various levels, i.e., patient level, image level, ROI level. 
We concatenate these predictions of the models introduced in Sec.~\ref{sec:task models} to form the vector $\fusionvec{\score{}{}}$, formally:
\begin{equation}
	\fusionvec{\score{}{}} = \score{\DensityModel}{} \cup \score{\FindingsModel}{v} \cup \score{\LocalizationModel}{v,n}
\end{equation}
\label{eq:fusion_predictions}

where $n$ is the number of considered detections per view. 
In case of no detected lesions by model $\LocalizationModel$ or less lesions than specified by $n$ are found, a probability of 0 is assigned, indicating that no (additional) lesions have been localized. 
For the malignancy prediction, only scores $\score{\LocalizationModel}{j,n}$ corresponding to malignant masses and calcifications are considered in the combined scores vector $\fusionvec{\score{}{}}$. In case no malignant lesions or less malignant lesions than specified by $n$ are found, a value of 0 is assigned.

\subsubsection{Fusion of Features ($\MetaModel{feat}$)} 

Apart from the fusion of prediction scores $\score{m}{}$, we also propose the fusion of feature vectors $\feature{m}{}, m \in \SetModels$ from the three different models. We extract features at the following stage in the networks:
\begin{itemize}
	\item $\feature{\DensityModel}{}$ is the 4096-dim., flattened, concatenated view-representations after global average pooling (see Fig. \ref{fig:density_patient_model})
	\item $\feature{\FindingsModel}{v}$ is the 1024-dim. representation for view image $I^v_i$, obtained after global average pooling (see Fig. \ref{fig:findings_model})
	\item $\feature{\LocalizationModel}{v,k}$ is the 1024-dim. representation for detection $k$ in $I^v_i$ (see Fig. \ref{fig:detection_model}).
\end{itemize}

\begin{figure}
	\centering
	\includegraphics[width=0.8\linewidth]{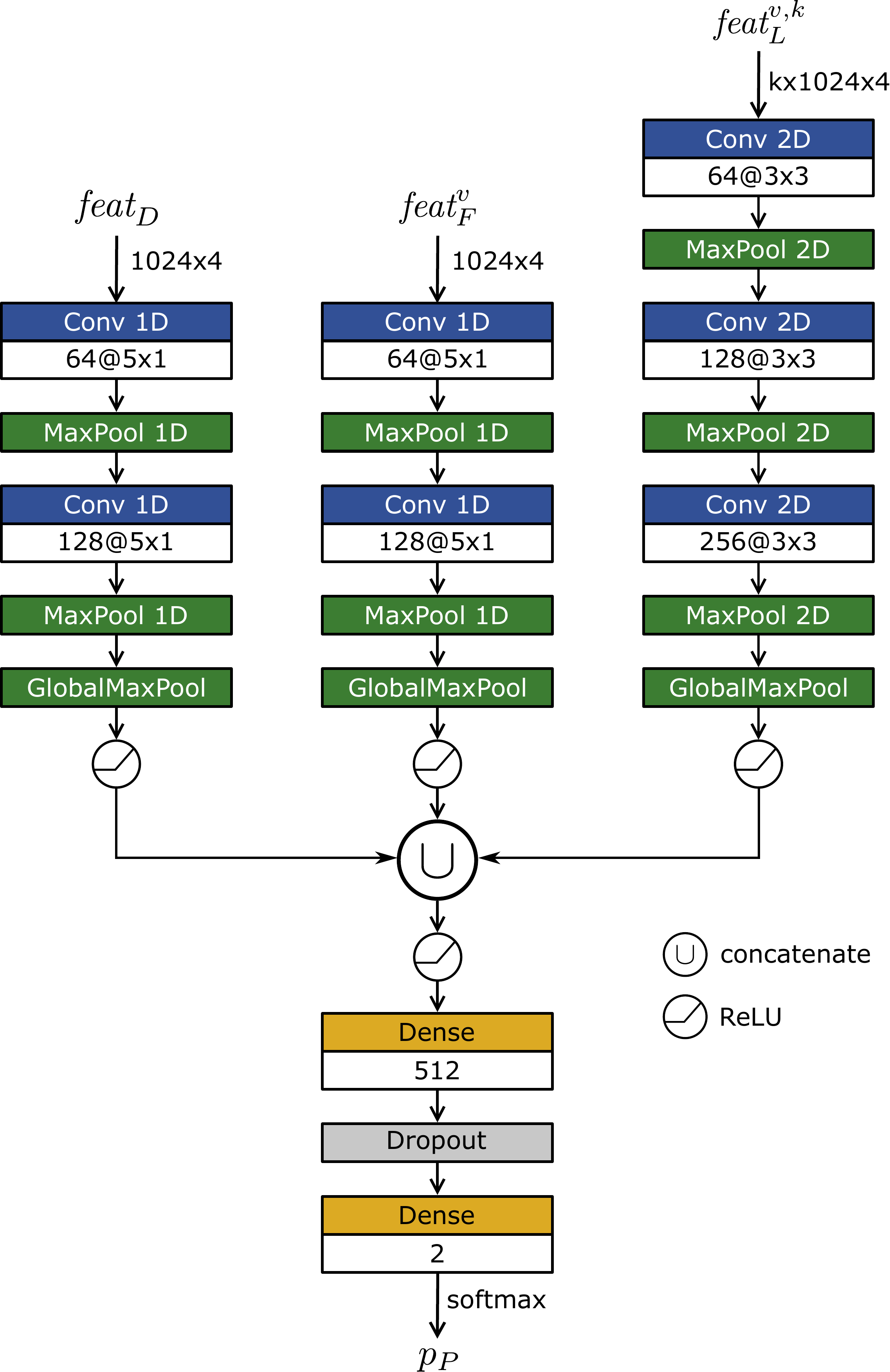}
	\caption{Patient meta-model $\MetaModel{feat}$}
	\label{fig:patient_meta_model}
\end{figure}

We propose an embedding network that takes the extracted, high-dimensional feature representations $\feature{m}{}$ as input in separate branches (see Fig.~\ref{fig:patient_meta_model}). Each channel corresponds to the respective features of a view image $I^v_i$. The density and findings branches consist of two convolution blocks, followed by pooling operations. The localization feature branch utilizes an additional convolution and pooling block for better feature learning. Before and after concatenation of all feature representations, we perform ReLU activations. The final classification part of the network consists of two dense layers with an intermediate dropout layer (dropout rate of 0.1) followed by a final softmax activation.

Again, we vary the number of lesions considered per view $n \in \{1,2,3,4,5\}$. In case no lesions are detected with model $\LocalizationModel$, or less lesions than specified by $n$, background features are pooled from the feature map and used as input. For the malignancy prediction, only features $\feature{\LocalizationModel}{j,n}$ corresponding to malignant masses and calcifications according to the localization model $\LocalizationModel$ are considered for the feature fusion. 
In case of no malignant lesions or less than specified by $n$, again background features are considered as model input. 


\section{Experimental Setup}
\label{sec:experimental setup}

We implemented our framework in Python, utilizing Keras~\cite{chollet2015short} with Tensorflow backend~\cite{tensorflowshort} for training the task-specific models $\DensityModel$, $\DensityModel^v$, $\LocalizationModel$, and patient meta-model $\MetaModel{feat}$. Additionally, we used the Tensorflow Object Detection API~\cite{objectdetectiontfshort} to train localization model $\LocalizationModel$ and scikit-learn for training patient meta-model $\MetaModel{score}$. Model training and experiments were conducted on an NVIDIA Titan X GPU (12 GB RAM).

\subsection{Training Details}
\label{sec:training details}

For the training of every task-specific model, we first segmented the breast with a basic, non-learning-based segmentation approach according to Shen et al.~\cite{shen2019}\footnote{\url{https://github.com/lishen/end2end-all-conv}}. Segmentation of the breast has been frequently used by related works as first preprocessing step, e.g., to clean/remove the background or for subsequent cropping to the breast area~\cite{shu2020,zhu2017,tardymateus2021,shen2019}. 
Similarly, we used the obtained breast mask to clean the background and for the sampling of patches inside the breast for pre-training the findings model $\FindingsModel$ (see Sec.~\ref{sec:training findings model}). The following set of random data augmentations was executed in each model training: horizontal flips, rotations (range: [-15,+15] degrees), and random sized crops (range: [85\%,100\%] of the image size). 
All image resizing operations were performed using bicubic resampling.

\subsubsection{Breast Density Model}
\label{sec:training breast density model}

All images were resized to 336~$\times$~224~$\times$~1 with rescaled intensities to the range [0,255] in floating-point precision to preserve the bit depth. 
Model training was conducted in a two-stage approach with Adam optimizer and cross-entropy loss: 
First, imagewise pre-training of the view model $\DensityModel^v$ (see Fig.~\ref{fig:density_view_model}) was performed for 25 epochs and an initial learning rate (lr) of 1e-3. Further, we employed Stochastic Weight Averaging (SWA) \cite{izmailov2018} with an initial epoch of 10 to increase the generalization capability of the model. In addition to the standard set of augmentations, random shears were conducted. 
Second, we trained the patient-wise model as shown in Fig.~\ref{fig:density_patient_model}. Each view branch was initialized with the SWA-weights from stage 1, and the complete model was trained for 25 epochs (lr = 1e-4). SWA was used with an initial epoch of 5. Horizontal flipping was not performed to preserve the original position of the breast in each view but instead blurring and grid distortion were additionally carried out. 
We reduced the learning rate by a factor of 0.2 with a patience of 5 epochs on the validation loss in both training stages.

\subsubsection{Findings Model}
\label{sec:training findings model}

\begin{figure}
	\centering
	\includegraphics[width=0.85\linewidth]{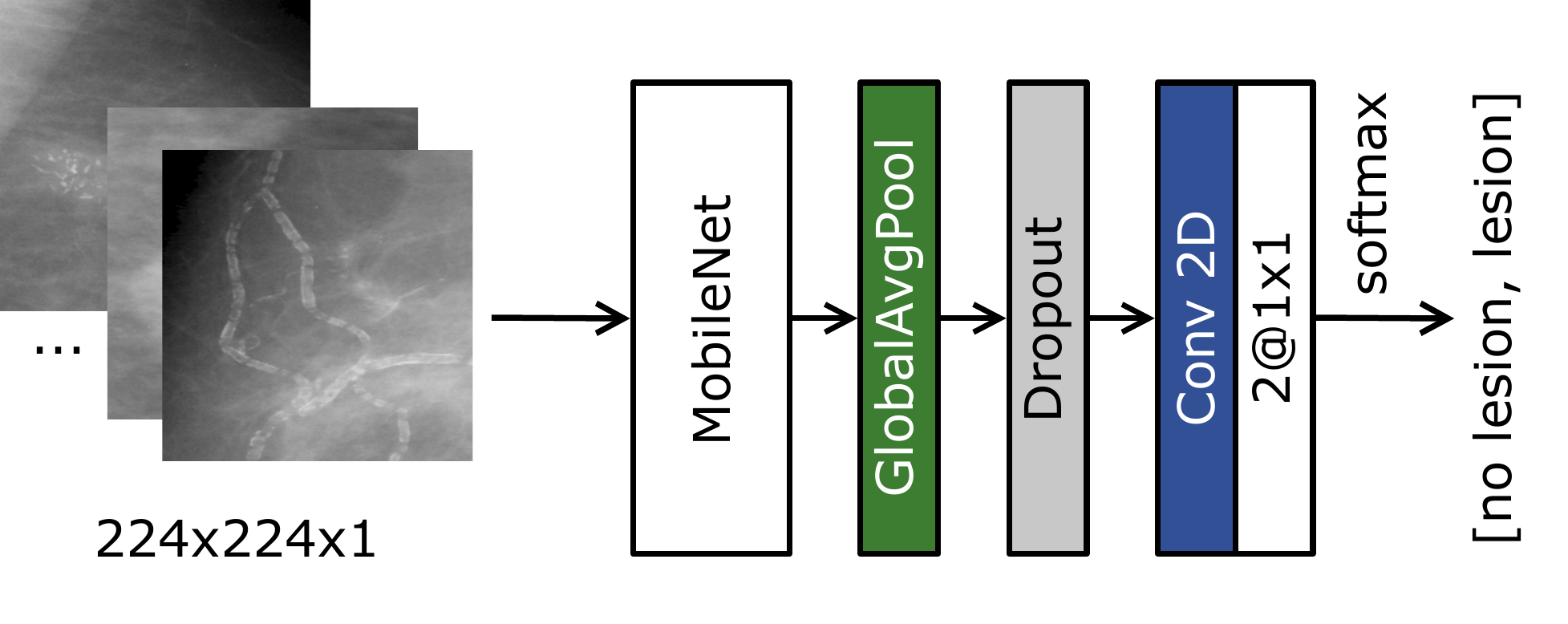}
	\caption{Patch Model}
	\label{fig:patch_model}
\end{figure}

We performed two-stage training of the findings model $\FindingsModel$, a strategy already successfully applied by recent works~\cite{shen2019,kim2020short,lotter2021short}. In both stages, the models were optimized using Adam with cross-entropy loss. First, a patch classifier (see Fig.~\ref{fig:patch_model}) was trained from scratch with patches of size 224~$\times$~224~$\times$~1, inspired by Shen et al.~\cite{shen2019}. We extracted an initial set by sampling 5 patches per lesion (overlap~$>$~90\% with lesion) and 5 patches from normal images (overlap~$>$~90\% with breast).
The patch model was trained with a batch size of 64 (lr~=~1e-4) and early stopping on the validation loss (patience = 10 epochs, tolerance~=~0.001). Additional augmentations were performed (vertical flips, transpose, and shift/scale/rotate) to further increase the diversity of patches. The model was fine-tuned in a second training iteration with a reduced learning rate of 1e-5.

In the second stage, we initialized the feature extractor of the findings model $\FindingsModel$ (see Fig.~\ref{fig:findings_model}) with the obtained patch weights. 
The full images were resized to 1152~$\times$~896~$\times$~1, rescaled to [0,1] and z-score normalized. 
$\FindingsModel$ was trained using a batch size of 6 (lr~=~1e-4). As opposed to the patch model, the validation AUC score was monitored as the criterion for early stopping (patience = 10 epochs, tolerance = 0.001). Additionally, SWA was used with an initial epoch of 5 which further improved the generalization capability. The model was fine-tuned in a second training round with lr~=~1e-5. 
In both training iterations of model $\FindingsModel$, vertical flips were additionally performed. In addition, stratified sampling was used to balance batches between images showing lesions and normal images~\cite{imblearn2017}.

\subsubsection{Localization Model}
\label{sec:training detection model}

The InceptionV2 backend was initialized with COCO-weights and then fine-tuned for the mammography lesion localization task for the four classes. The ground truth bounding boxes required to train the Faster R-CNN model were derived from the pixelwise annotated lesions. We consider the axis-aligned minimum bounding box which encloses the lesion. The model was trained according to the pipeline split, whereby only images with at least one lesion were considered for training. We resized the view images to 2700 $\times$ 1200 and trained $\LocalizationModel$ with SGD (momentum~=~0.9, lr~=~1e-4) for 100k iterations and a batch size of 2. In addition to the default data augmentation strategies, bounding boxes were randomly jittered with a ratio of 0.005.

\subsubsection{Patient Meta-Model}

We performed a parameter search over the number of considered lesions $n \in \{1,2,3,4,5\}$ for $\MetaModel{score}$ and $\MetaModel{feat}$ and trained all models according to the predefined data split for the lesion and malignancy prediction. Best models were selected based on validation AUC and recall.

\paragraph{Prediction score fusion}

Prediction scores were concatenated according to Eq.~\ref{eq:fusion_predictions} to obtain one feature vector $\fusionvec{\score{}{}}$ per patient. We varied the number of detected lesions $n \in \{1,2,3,4,5\}$ considered per view and included only their scores. For comparison, a classic SVM with RBF kernel, a multilayer perceptron (MLP), and a random forest were trained. Parameter search was performed over the parameters of the individual models and selected the model with highest validation AUC: SVM RBF (C = \{1e-1, 1e-2, 1e-3, 1e-4, 1, 10, 100, 500, 1000\}), random forest (number of trees = \{3,5,7,10,15,20\}), MLP (layer configuration = \{[$\lvert \fusionvec{\score{}{}} \rvert$, 2], [$\lvert \fusionvec{\score{}{}} \rvert$, $\lvert \fusionvec{\score{}{}} \rvert$, 2], [$\lvert \fusionvec{\score{}{}} \rvert$, $\lvert \fusionvec{\score{}{}} \rvert$/2, 2]\}) .

\paragraph{Feature fusion}

Before feeding the feature representations to $\MetaModel{feat}$, they were normalized with $\phi$, where $\phi~:~\mathds{R}^n~\mapsto~[-1,1]$, resulting in normalized representations $\phi(\feature{\DensityModel}{})$, $\phi(\feature{\FindingsModel}{j})$, and $\phi(\feature{\LocalizationModel}{j,k})$. 
We optimized $\MetaModel{feat}$ with Adam using cross-entropy loss and a batch size of 8 and lr~=~5e-4. Early stopping was used with a patience of 10 epochs on the validation loss (tolerance = 0.001). Again, batches were balanced to ensure equal distribution of classes.

\subsection{Evaluation Metrics}
\label{sec:metrics}

We compare the performance of classification-related tasks $\DensityModel$ and $\FindingsModel$ by calculating widely used metrics in the field: the true positive rate (TPR), also referred to as \textit{sensitivity} or \textit{recall}, the true negative rate (TNR), also referred to as \textit{specificity}, accuracy, and F1-score (F1), i.e., the harmonic mean of precision and recall. Further, we calculate the area under the ROC curve (AUC), which shows the TPR against the false-positive rate (1 - specificity). Additionally, we provide the area under the precision-recall curve (AUPRC) for comparisons with recent studies~\cite{shachor2020,kyono2019,shengeras2021short}. 
For the localization model $\LocalizationModel$, we provide FROC curves to measure its detection performance and calculate the number of false positives per image (FPI) at given TPR rates.


\section{Results and Discussion}
\label{sec:results}

This section summarizes intermediate results obtained with task-specific models (see Section~\ref{sec:results individual models}) as well as final predictions obtained via score and feature fusion (see Section~\ref{sec:results meta models}). Section~\ref{sec:ablation studies} summarizes ablation study results, and finally, Section~\ref{sec:discussion} provides an in-depth discussion and analysis of the presented results. 

We performed Wilcoxon signed-rank tests on the predictions for task-specific models, fusion models, as well as for the performed ablation studies. Similar to recent studies~\cite{shu2020,tardymateus2021,rodriguezruiz2019}, we set the significance level to $\alpha = 0.05$.

\subsection{Performance on Individual Tasks}
\label{sec:results individual models}

\begin{figure}
	\centering
	\subfloat[breast density $\DensityModel$]{\includegraphics[width=4.5cm]{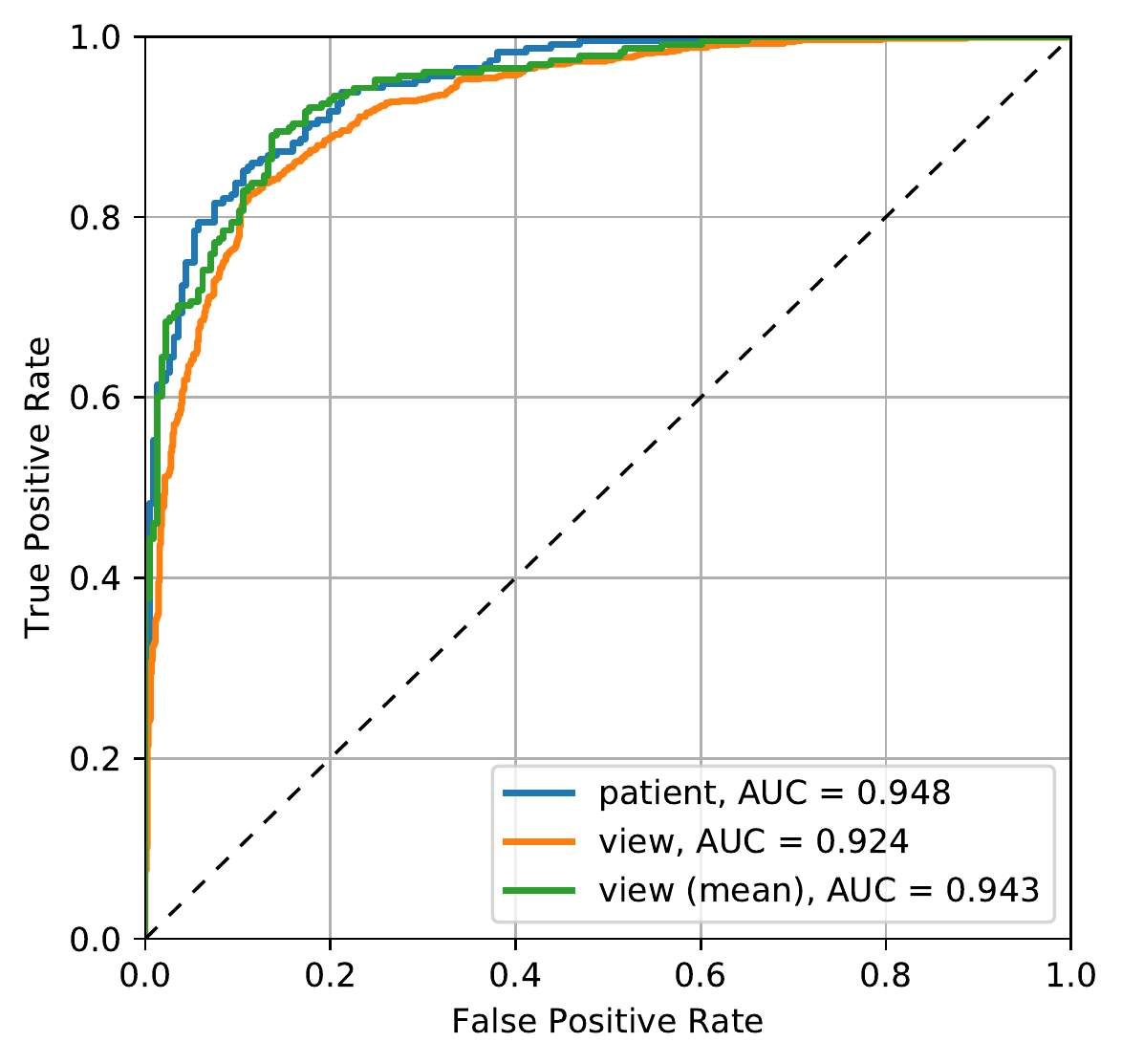}%
		\label{fig:density_ROC}}
	\subfloat[lesion localization $\LocalizationModel$]{\includegraphics[width=4.5cm]{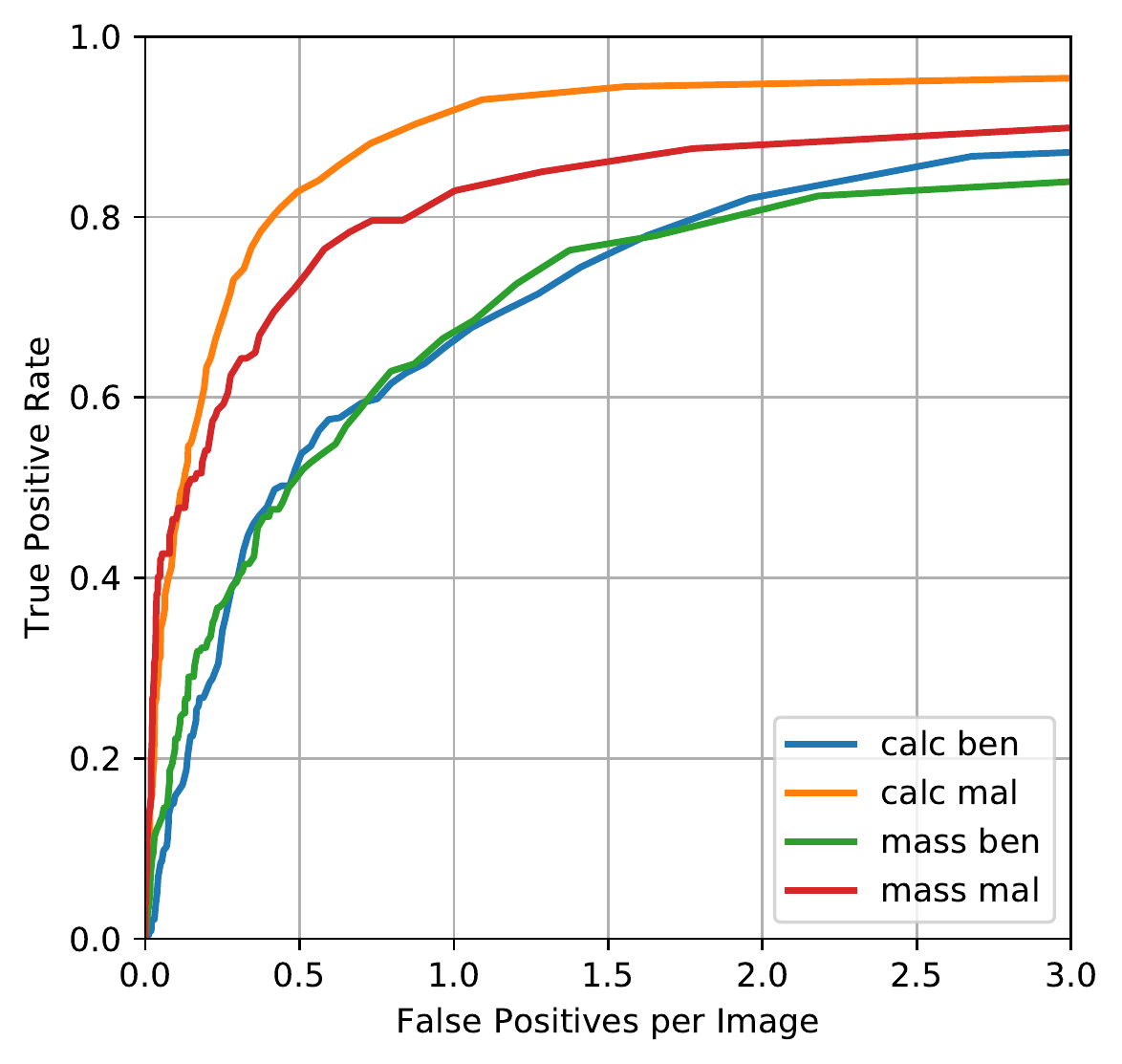}%
		\label{fig:detection_FROC_4cls}}
	\hfil
	\caption{ROC and FROC curves of individual models.}
	\label{fig:individual model results}
\end{figure}

Task-specific model results are visualized in terms of ROC and FROC curves in Fig.~\ref{fig:individual model results}.

\subsubsection{Breast Density Classification}
\label{sec:results breast density}

We report an AUC score of 0.948 of density model $\DensityModel$ on the test set with TPR~=~0.882 and specificity~=~0.832 (F1 = 0.861). As depicted in Fig.~\ref{fig:density_ROC}, the final model $\DensityModel$ on patient level (blue) shows a minor improvement in terms of AUC (AUC~=~0.943, $p < 0.001$) compared to the aggregated predictions $\text{mean}(\DensityModel^v)$ of the imagewise model $\DensityModel^v, v \in$ \{L-CC, L-MLO, R-CC, R-MLO\} on patient level (TPR~=~0.833, specificity = 0.889, F1 = 0.858). Further, we observe a significantly higher sensitivity with $\DensityModel^v$ compared to $\text{mean}(\DensityModel^v)$ ($p < 0.001$) at  similar accuracies (see Table~\ref{tab:breast_density_results_comparison}).
On image level, we report an AUC of 0.924 with $\DensityModel^v$ (TPR~=~0.815, specificity = 0.894, F1 = 0.849, accuracy~=~0.854).

Table~\ref{tab:breast_density_results_comparison} summarizes density classification results reported in the literature. We report higher accuracy scores on DDSM compared to Oliver et al.~\cite{oliver2008}, who tested only on a subset of 831 R-MLO images, while our method was evaluated on 453 patients, i.e., 1812 view images. While our model performs slightly beneath published works, these methods were trained utilizing significantly larger datasets, e.g., the dataset by Wu et al.~\cite{wu2018short} comprises 200k exams (80\% train / 20\% test data).

\begin{table}[]
	\centering
	\caption{Overview on reported density classification accuracies (acc.) in related works and obtained with our model $\DensityModel$. Methods indicated with * use one image as input, those without utilize all four view images.}
	\label{tab:breast_density_results_comparison}
	\begin{tabular}{|l|c|c|c|}
		\hline
		Method 									& Data 		&  Acc. (4 cls)	& Acc. (2 cls) \\
		\hline \hline
		Wu~\cite{wu2018short}			 		& private (NYU)	& 0.767 & 0.865 (derived) \\
		\hline
		Lehman~\cite{lehman2019} *	& private & 0.770 & 0.870 (derived) \\
		\hline
		Kaiser~\cite{kaiser2019} 	& private & --		& 0.881 \\
		\hline
		Oliver~\cite{oliver2008} *	& DDSM (R-MLO) & 0.772  & 0.842 \\
 		\hline
		Ours * ($\DensityModel^v$)	& DDSM			& --	& 0.854 \\
		\hline
		Ours ($\text{mean}(\DensityModel^v)$) & DDSM & --	& 0.861 \\
		\hline
		Ours ($\DensityModel$)		& DDSM			& --	& 0.857	\\
		\hline
	\end{tabular}
\end{table}

\subsubsection{Findings Classification}
\label{sec:results findings}

For the task of classifying images into those with any lesion and those without, model $\FindingsModel$ reaches an AUC score of 0.921 on test data with TPR~=~0.881 and specificity~=~0.802 (F1~=~0.878). 

To the best of our knowledge, there is only the work by Lotter et al.~\cite{lotter2017}, who used the presence/absence of lesions as classification target for pre-training their model on patch-level, thus, they did not report performance measures on image level.

\subsubsection{Lesion Localization}
\label{sec:results lesion detection}

We report TPR rates of 0.84 for malignant masses, 0.93 for malignant calcifications, 0.70 for benign masses, and 0.68 for benign calcifications by localization model $\LocalizationModel$ on test images with lesions, as summarized in Table~\ref{tab:lesion_localization_results_comparison}. Fig.~\ref{fig:detection_FROC_4cls} shows corresponding FROC curves. A lesion is considered detected if the intersection over union (IoU) of the detected bounding box with the ground truth bounding box is $\geq$ 0.2, or if the center of the detected bounding box lies within the ground truth bounding box~\cite{ribli2018}. 
On normal images in the test set (105 patients, i.e., 420 view images), we detect 386 false-positive lesions in 188/420 images. On the 348 abnormal cases, we detect 2478 false-positive lesions.

\begin{figure}
	\centering
	\subfloat{\includegraphics[height=7cm]{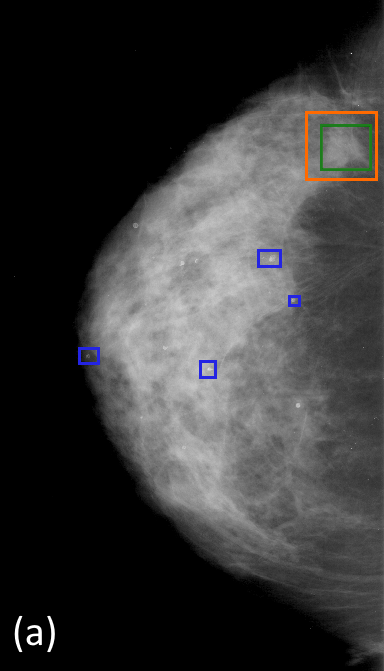}%
		\label{fig:loclization_correct+wrong}}
	\hfil
	\subfloat{\includegraphics[height=7cm]{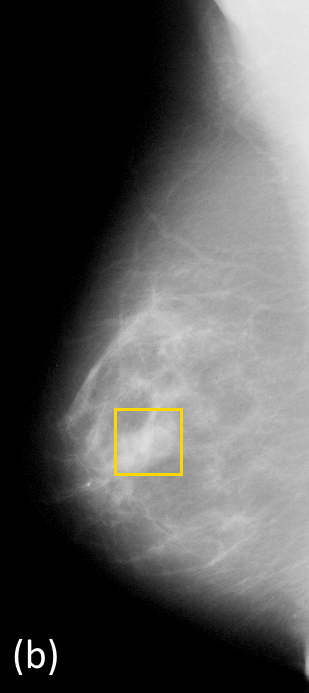}%
		\label{fig:localization_wrong_mass}}
	\caption{(a) R-CC image with correctly localized malignant mass (green = ground truth, orange = detected) and additional detected benign calcifications (blue) not present in ground truth, (b) R-MLO image with false-positive benign mass (yellow). Best viewed in color.}
	\label{fig:visual_localization_results}
\end{figure}

Fig.~\ref{fig:visual_localization_results} shows visual samples of correct and false-positive detected lesions. Overall, we report lower detection rates for benign lesions compared to malignant lesions, a phenomenon also observed in the literature~\cite{agarwal2020}. 
As visible in Fig.~\ref{fig:visual_localization_results}, one reason for the lower performance of model $\LocalizationModel$ is the detection of small calcifications (in blue), which appear very similar to benign calcifications but are not annotated as such in the ground truth. Another aspect is the misclassification of denser breast tissue with masses as well as overlaps of benign and malignant masses that can occur due to non-maxima suppression performed on class-level.   

\begin{table}[]
	\centering
	\caption{Overview on lesion localization results reported in related works and results obtained with our model $\LocalizationModel$ (OMI-H = OPTIMAM database, Hologic scanner images only, *~=~subset of 300 images used).}
	\label{tab:lesion_localization_results_comparison}
	\begin{tabular}{|l|c|c|c|}
		\hline
		Method 									& Train/Test Data 		& Lesion	& TPR @ FPI \\
		\hline \hline
		Agarwal~\cite{agarwal2020}		 		& \parbox{2cm}{OMI-H / OMI-H } & mass & 0.93 @ 0.78  \\
										 		& \parbox{2.1cm}{OMI-H / INbreast} & mal. mass & 0.99 @ 1.17 \\
												&  & ben. mass & 0.85 @ 1.0 \\
		\hline
		Ribli~\cite{ribli2018}					& \parbox{2cm}{DDSM, private / INbreast} & mal. lesion & 0.9 @ 0.3 \\	
		\hline
		Akselrod-B.~\cite{akselrodballin2017}	& \parbox{2cm}{private /\\ INBreast, private} & mass	& 0.90 @ 0.3 \\
		\hline
		Anitha~\cite{anitha2017}			& -- / DDSM* & mass & 0.925 @ 1.06 \\
		\hline
		Ours								& DDSM / DDSM	& mal. mass & 0.84 @ 1.0 \\
											& 	& mal. calc. & 0.93 @ 1.09 \\ 
											& 	& ben. mass & 0.70 @ 1.06 \\
											& 	& ben. calc. & 0.68 @ 1.06 \\
		\hline
	\end{tabular}
\end{table}

Table~\ref{tab:lesion_localization_results_comparison} provides an overview on localization results reported in the literature. However, the localization performances of the different methods cannot be compared directly due to the large differences in the datasets and varying criteria for correctly detected lesions. The method by Agarwal et al.~\cite{agarwal2020}, for example, utilizes the much larger OPTIMAM database, while Anitha et al.~\cite{anitha2017}, on the other hand, use only a subset of the DDSM set rather than the full dataset.

\subsection{Patient Meta-Model Results}
\label{sec:results meta models}

\begin{figure}
	\centering
	\subfloat{\includegraphics[width=4.6cm]{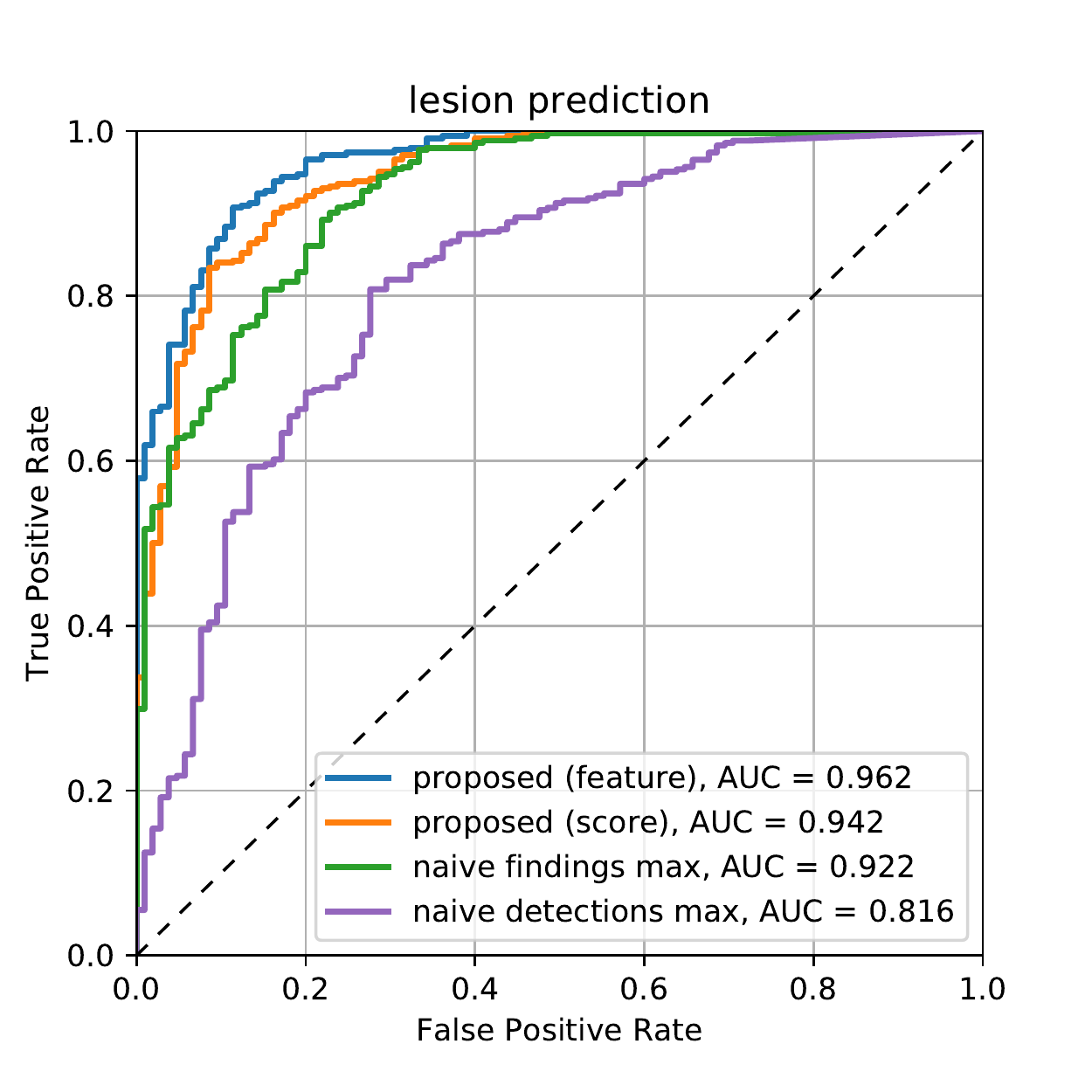}%
		\label{fig:lesion_models_ROC}}
	\subfloat{\includegraphics[width=4.6cm]{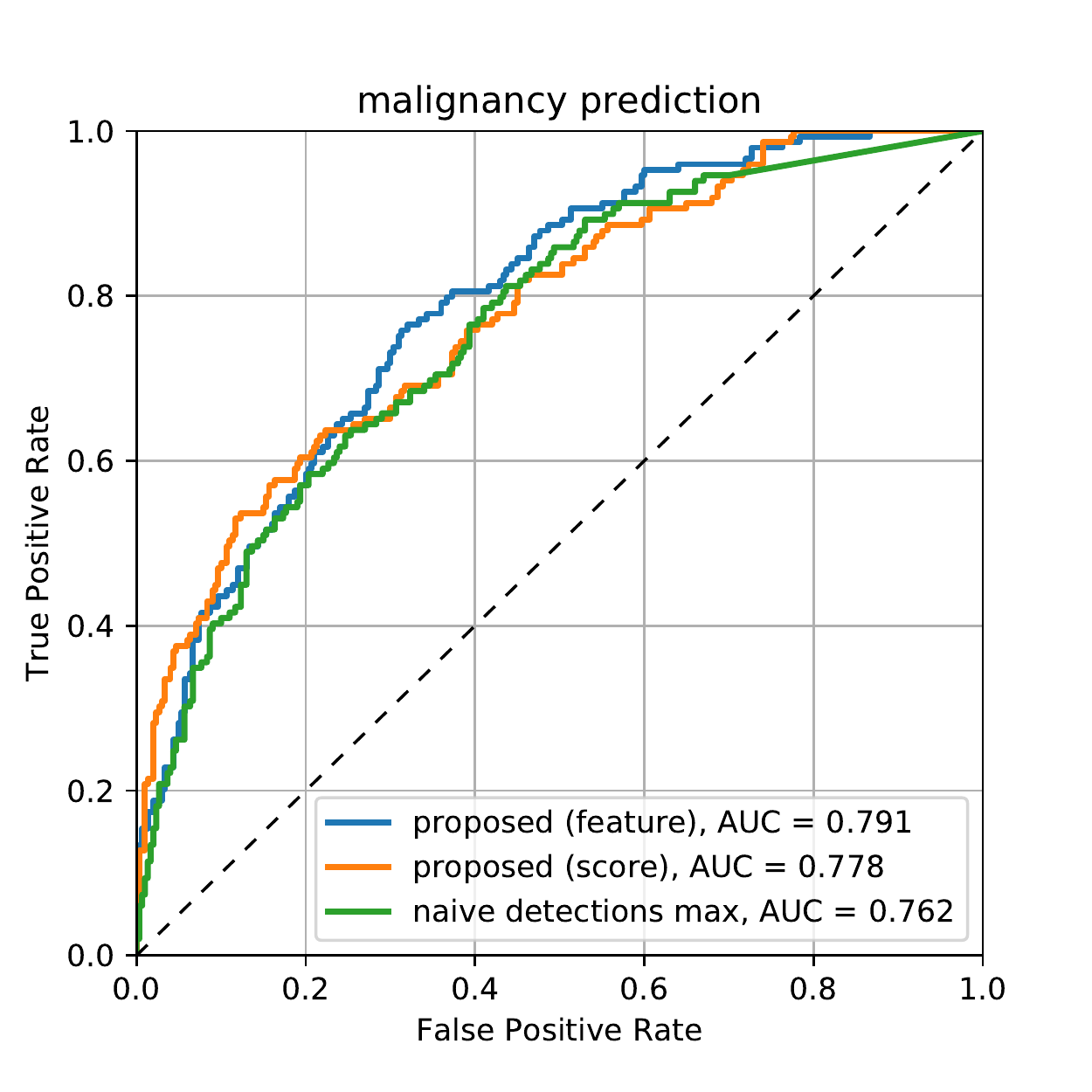}%
		\label{fig:malignancy_models_ROC}}
	\caption{ROC curves of patient models $\MetaModel{score}$ and $\MetaModel{feat}$ for the lesion prediction (left), and malignancy prediction (right).}
	\label{fig:meta_models_ROC}
\end{figure}

ROC curves for feature and score fusion $\MetaModel{feat}$ and $\MetaModel{score}$ (with MLPs), respectively, for both patient predictions, are shown in Fig.~\ref{fig:meta_models_ROC}. Table~\ref{tab:meta_model_results} summarizes quantitative performance measures on test data. We additionally trained our fusion models without density information (indicated with * in Table~\ref{tab:meta_model_results}) and compared all our fusion results to standard ensembling, i.e., taking the maximum of prediction scores. A detailed statistical significance analysis for all fusion models is summarized in Table~\ref{tab:p_value_analysis}.

\begin{table}[]
	\centering
	\caption{Performance metrics of patient fusion models $\MetaModel{score}$ (MLPs) and $\MetaModel{feat}$ on test data. Models marked with * indicate exclusion of breast density information. max($\score{\FindingsModel}{v}$) and max($\score{\LocalizationModel}{v}$) denote the naive score maximum of findings model $\FindingsModel$ and detection model $\LocalizationModel$ for lesion prediction and malignancy prediction, respectively.}
	\label{tab:meta_model_results}
	\begin{tabular}{|l|c|c|c|c|c|c|}
		\hline
		Model 								& Target	& AUC 	& F1	& TPR	& Specificity \\
		\hline \hline
		$\MetaModel{score}$ 				& lesion	& 0.942	& 0.932	& 0.933	& 0.771	\\
		\hline
		$\MetaModel{score}$* 	 			& lesion	& 0.941	& 0.928	& 0.919	& 0.800 \\
		\hline 
		max($\score{\FindingsModel}{v}$)	& lesion	& 0.922	& 0.938	& 0.974	& 0.667 \\
		\hline \hline
		$\MetaModel{feat}$ 					& lesion  	& 0.962	& 0.948	& 0.956	& 0.800 \\
		\hline
		$\MetaModel{feat}$*					& lesion	& 0.959	& 0.943	& 0.939 & 0.829 \\
		\hline
		max($\score{\FindingsModel}{v}$)	& lesion	& 0.922	& 0.938	& 0.974	& 0.667 \\
		\hline \hline
		$\MetaModel{score}$ 				& malignancy	& 0.778	& 0.601	& 0.591 & 0.813 \\
		\hline
		$\MetaModel{score}$* 				& malignancy	& 0.774	& 0.523	& 0.578 & 0.857 \\
		\hline
		max($\score{\LocalizationModel}{v}$)& malignancy	& 0.762 & 0.578	& 0.570	& 0.800 \\
		\hline \hline
		$\MetaModel{feat}$ 					& malignancy	& 0.791	& 0.603	& 0.638	& 0.763 \\
		\hline
		$\MetaModel{feat}$*					& malignancy	& 0.789	& 0.581	& 0.577 & 0.797 \\ 
		\hline
		max($\score{\LocalizationModel}{v}$) & malignancy	& 0.762 & 0.578	& 0.570	& 0.800 \\
		\hline
	\end{tabular}
\end{table}

For $\MetaModel{score}$, we obtain the best results in terms of AUC and TPR with MLPs for both patient predictions, compared to SVMs and random forests (see Table~\ref{tab:score_fusion_results}). In terms of the number of included lesions $n$ in the meta-models, the best results reported in Table~\ref{tab:meta_model_results} and Table~\ref{tab:score_fusion_results} are obtained with $n=3$ for the lesion prediction ($\MetaModel{score}$ and $\MetaModel{feat}$), and $n=3$ ($\MetaModel{score}$) and $n=1$ ($\MetaModel{feat}$) for the malignancy prediction. A detailed overview of quantitative results for $\MetaModel{score}$ (SVM, random forest, MLP) and $\MetaModel{feat}$ for different numbers of included lesions $n$ is provided in the supplemental material. 

\begin{table}[]
	\centering
	\caption{Comparison of performance metrics of $\MetaModel{score}$ for MLP, SVM, and random forest for both patient predictions.}
	\label{tab:score_fusion_results}
	\begin{tabular}{|l|c|c|c|c|c|c|}
		\hline
		Model 				& Target		& AUC 	& F1	& TPR	& Specificity \\
		\hline \hline
		MLP 				& lesion		& 0.942	& 0.932	& 0.933	& 0.771	\\
		\hline
		SVM 				& lesion		& 0.935	& 0.928	& 0.916	& 0.810	\\
		\hline
		Random Forest		& lesion		& 0.929	& 0.924	& 0.919	& 0.771	\\
		\hline \hline
		MLP 				& malignancy	& 0.778	& 0.601	& 0.591 & 0.813 \\
		\hline
		SVM 				& malignancy	& 0.763	& 0.552	& 0.483 & 0.867 \\
		\hline
		Random Forest		& malignancy	& 0.776	& 0.581	& 0.564 & 0.813 \\
		\hline
	\end{tabular}
\end{table}

Overall, we report an increase in terms of AUC between 0.02 and 0.04 for the lesion prediction with score fusion and feature fusion, respectively, when comparing to the naive score maximum across the four views max($\score{\FindingsModel}{v}$) ($p < 0.001$ for both). A slightly smaller increase is obtained for the malignancy prediction, ranging from 0.016 ($\MetaModel{score}$) to 0.029 ($\MetaModel{feat}$), as compared to max($\score{\LocalizationModel}{v}$) ($p < 0.001$ for both). 
We report higher AUC scores and increased sensitivity, i.e., TPR, with feature fusion models compared to score fusion models for both patient predictions ($p < 0.001$). However, for the malignancy prediction, we observe a reduced specificity for feature fusion as compared to score fusion.

\begin{table*}[]
	\centering
	\caption{Statistical significance analysis for fusion models $\MetaModel{score}$ and $\MetaModel{feat}$ with p-values $p < 0.05$ denoting statistically significance (bold font). Models marked with * indicate exclusion of breast density information. max($\score{\FindingsModel}{v}$) and max($\score{\LocalizationModel}{v}$) denote the score maximum of findings model $\FindingsModel$ and detection model $\LocalizationModel$ for lesion and malignancy prediction, respectively. p-values are given for the lesion prediction and the malignancy prediction (separated by "/").}
	\label{tab:p_value_analysis}
	\begin{tabular}{|l|c|c|c|c|c|c|c|}
		\hline
		Model 								& $\MetaModel{score}$	& $\MetaModel{score}$*	& $\MetaModel{feat}$	& $\MetaModel{feat}$*	& max($\score{\FindingsModel}{v}$) 		& 		max($\score{\LocalizationModel}{v}$) \\
		\hline
		$\MetaModel{score}$ 				& $\infty$	& \textbf{$<$ 0.001} / \textbf{$<$ 0.001}	& \textbf{$<$ 0.001} / \textbf{0.023}	& \textbf{$<$ 0.001} / \textbf{0.002} 	& \textbf{$<$ 0.001} / --- & \textbf{$<$ 0.001} / \textbf{$<$ 0.001} \\
		\hline
		$\MetaModel{score}$* 				& \textbf{$<$ 0.001} / \textbf{$<$ 0.001} & $\infty$	& \textbf{$<$ 0.001} / \textbf{$<$ 0.001}	& \textbf{$<$ 0.001} /	0.237 & \textbf{$<$ 0.001} / --- & \textbf{$<$ 0.001} / 0.999 \\
		\hline 
		$\MetaModel{feat}$ 					& \textbf{$<$ 0.001} / \textbf{0.023} & \textbf{$<$ 0.001} / \textbf{$<$ 0.001}	& $\infty$	&  \textbf{$<$ 0.001} / \textbf{$<$ 0.001} & \textbf{$<$ 0.001} / --- & \textbf{$<$ 0.001} / \textbf{$<$ 0.001} \\
		\hline
		$\MetaModel{feat}$*					& \textbf{$<$ 0.001} / \textbf{0.002} & \textbf{$<$ 0.001} / 0.237	& \textbf{$<$ 0.001} / \textbf{$<$ 0.001}	& $\infty$  & \textbf{$<$ 0.001} / --- & \textbf{$<$ 0.001} / 0.920 \\
		\hline
		max($\score{\FindingsModel}{v}$)	& \textbf{$<$ 0.001} / --- & \textbf{$<$ 0.001} / ---	& \textbf{$<$ 0.001} / --- & \textbf{$<$ 0.001} / ---	& $\infty$ & \textbf{$<$ 0.001} / --- \\
		\hline 
		max($\score{\LocalizationModel}{v}$) & \textbf{$<$ 0.001} / \textbf{$<$ 0.001} & \textbf{$<$ 0.001} / 0.999 & \textbf{$<$ 0.001} / \textbf{$<$ 0.001} & \textbf{$<$ 0.001} / 0.920 & \textbf{$<$ 0.001} / --- & $\infty$ \\
		\hline
	\end{tabular}
\end{table*}

\begin{figure*}
	\centering
	\includegraphics[width=\linewidth]{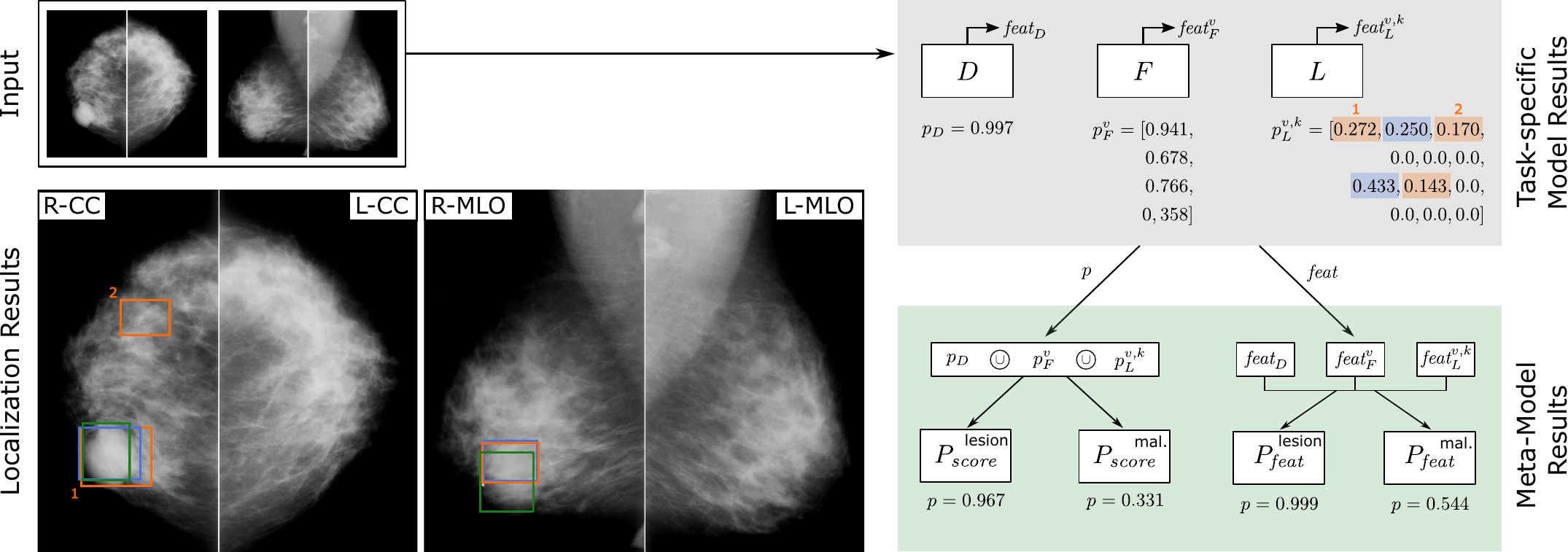}%
	\caption{Illustration of the benefits of our mammography pipeline: The patient has a malignant mass (green = ground truth) in the right breast. Model $\LocalizationModel$ was able to localize the malignant mass (orange), but with low confidence that is not sufficient to be reliably counted as detection. Low confidence localizations were also found by $\LocalizationModel$ for an additional malignant mass and a benign mass (blue). Detection scores $\score{L}{v,k}$ are colored according to their class label as detected by $\LocalizationModel$. Results show that both fusion models $\MetaModel{feat}$ are able to circumvent the low scores and correctly classify the patient (malignancy~=~0.544, lesion~=~0.999). Best viewed in color.}
	\label{fig:result_pipeline}
\end{figure*}

Fig.~\ref{fig:result_pipeline} shows a sample result obtained with our proposed mammography pipeline: 
The localization model $\LocalizationModel$ was able to correctly localize the malignant mass in the right breast but also falsely detected a benign mass at the same location. In R-MLO view, the benign detection was given a higher confidence than the malignant detection, which would lead to false patient-level results in case we solely rely on $\LocalizationModel$. However, the feature fusion models $\MetaModel{feat}$ were able to correctly classify the patient in terms of the lesion and malignancy prediction, while $\MetaModel{score}$ failed for the malignancy prediction (decision threshold~=~0.5).

\subsection{Ablation Studies}
\label{sec:ablation studies}

Complementary to the training setup described in Section~\ref{sec:training details}, we performed additional experiments to support our pre-training strategies for task-specific models $\DensityModel$ and $\FindingsModel$. Further, we investigated the influence of breast density information in the fusion models.

\subsubsection{Pre-training of $\DensityModel$}

We retrained the density model $\DensityModel$ without pre-training the view model $\DensityModel^v$ with the same training parameters (see Section~\ref{sec:training breast density model}) except for a lower learning rate of 1e-3. We obtain a significantly lower AUC score of 0.900 ($p < 0.001$) with this model. Further we report TPR~=~0.934, specificity = 0.690, F1 = 0.817, and a significantly lower accuracy of 0.797 as compared to $\DensityModel$ ($p < 0.001$).

\subsubsection{Pre-training of $\FindingsModel$}

Further, we retrained the findings classifier $\FindingsModel$ without patch-wise pre-training with the same training parameters as described in Section~\ref{sec:training findings model}. The model without pre-training achieves a significantly lower AUC score of 0.895 ($p < 0.001$) and sensitivity~=~0.816, F1~=~0.846, specificity~=~0.817.

\subsubsection{Breast Density Ablation Study}
\label{sec:breast density ablation study results}

As breast density is an essential risk factor for breast cancer~\cite{destounis2020}, we retrained our patient meta-models with the same training parameters but excluded breast density features and scores for $\MetaModel{feat}$ and $\MetaModel{score}$, and denoted the obtained models $\MetaModel{feat}$* and $\MetaModel{score}$*, respectively. 
Results in Table~\ref{tab:meta_model_results} show higher AUC scores and a higher TPR for all fusion models $\MetaModel{score}$ and $\MetaModel{feat}$ when \textit{including} breast density information ($p < 0.001$, as summarized in Table~\ref{tab:p_value_analysis}). No statistically significant difference can be reported when comparing $\MetaModel{score}$* and $\MetaModel{feat}$* with max($\score{\LocalizationModel}{v}$) for the malignancy prediction with p-values $p=0.999$ and $p=0.920$, respectively. Further, no significant difference can be observed between $\MetaModel{score}$* and $\MetaModel{feat}$* for the malignancy prediction ($p=0.237$). These results indicate that the inclusion of breast density can yield improved classification performance.

\subsection{Discussion}
\label{sec:discussion}

\subsubsection{Breast Density}
\label{sec:breast density discussion}

\begin{table}[]
	\centering
	\caption{Comparison of different measures obtained with $\DensityModel$ and $\text{mean}(\DensityModel^v)$ at various decision thresholds.}
	\label{tab:breast_density_decision_threshold}
	\begin{tabular}{|l|c|c|c|c|c|}
		\hline
		Model							& Threshold	& TPR 	& F1	& Spec.	&  Acc. (2 cls) \\
		\hline \hline
		$\DensityModel$					& 0.6		& 0.860	& 0.871	& 0.885	& 0.872 \\
		\hline
		$\text{mean}(\DensityModel^v)$	& 0.6		& 0.781	& 0.838	& 0.916	& 0.848 \\
		\hline 
		$\DensityModel$					& 0.7		& 0.816	& 0.859	& 0.916	& 0.866 \\
		\hline	
		$\text{mean}(\DensityModel^v)$	& 0.7		& 0.711	& 0.804 & 0.942 & 0.826 \\
		\hline
		$\DensityModel$					& 0.8		& 0.763	& 0.841 & 0.947	& 0.855 \\
		\hline
		$\text{mean}(\DensityModel^v)$	& 0.8		& 0.684	& 0.800	& 0.973	& 0.828 \\
		\hline
	\end{tabular}
\end{table}

We investigated the patient density model $\DensityModel$ and aggregated view model $\text{mean}(\DensityModel^v)$ in depth and varied the decision threshold (see Table~\ref{tab:breast_density_decision_threshold}). 
The results show that model $\DensityModel$ yields more reliable predictions with high confidence, and thus, higher sensitivities, accuracies, and F1 scores at various thresholds compared to the aggregated view model ($p < 0.001$). 
Such automated tools that deliver trustworthy, reproducible measures are of increasing importance in clinical practice, especially for breast density assessment where subjectivity and high inter-observer variability are well-known issues~\cite{sprague2016short,kaiser2019,destounis2020}. As breast density is considered an important risk factor for the development of breast cancer, reliability and reproducibility are key aspects when it comes to standardized density reporting which may trigger supplemental/personalized screening procedures~\cite{sprague2016short,conant2018,destounis2020}.

\subsubsection{Comparison to Related Work}

\begin{table*}[t!]
	\centering
	\caption{Results obtained with fusion models $\MetaModel{feat}$ compared to classification results reported in related works. Train and test datasets utilized by the respective methods are separated with "/".}
	\label{tab:final_results_comparison}
	\begin{tabular}{|l|c|c|c|c|c|c|}
		\hline
		Method 									& Train/Test Data 			& Fusion-Level 	& Result-Level 	& Target	& AUC 	& AUPRC\\
		\hline \hline
		Shen~\cite{shen2019} 					& CBIS-DDSM	/ CBIS-DDSM				& -		& image			& mal. 		& 0.87	& - \\
		& CBIS-DDSM + INbreast / INbreast	& -		& image			& mal.	 	& 0.95	& - \\
		\hline
		Shu~\cite{shu2020} 						& CBIS-DDSM	/ CBIS-DDSM				& -		& image			& mal.		& 0.838	& - \\
		& INbreast / INbreast				& -		& image 		& mal.		& 0.934	& - \\
		\hline
		Ribli~\cite{ribli2018}					& DDSM, private / INbreast	& -				& \parbox{1.7cm}{\centering breast (max+avg)}	& mal.		& 0.95	& - \\
		\hline
		Lotter~\cite{lotter2021short}			& \parbox{3.8cm}{\centering DDSM, OPTIMAM, private /\\ OPTIMAM}	& ROI	& \parbox{1.7cm}{\centering patient (avg+max)} & mal.	& 0.963 $\pm$ 0.003 & - \\
		\hline
		Kooi~\cite{kooi2017}					& private / private (NL screening) & ROI	& image			& mal. mass & 0.941	& - \\
		\hline
		Shen~\cite{shengeras2021short}			& CBIS-DDSM / CBIS-DDSM				& image	& breast (avg)	& mal.		& 0.833	& - \\
		& private / private	(NYU)				& image	& breast (avg)	& mal.		& 0.891 & 0.390 \\
		\hline
		Shachor~\cite{shachor2020} 				& DDSM / DDSM	 	& ROIs (from CC+MLO)	& breast	 	&	ben./mal. calc. & 0.661 & -\\
		\hline
		Kyono~\cite{kyono2019}					& private / private (Tommy trial)	& patient 	& patient	& mal.		& 0.824 $\pm$ 0.016 & 0.580 $\pm$ 0.028\\
		\hline
		Ours ($\MetaModel{feat}$)				& DDSM / DDSM			  			& patient	& patient 	& mal.		& 0.791 & 0.660 \\
		&			 						& patient	& patient 	& lesion	& 0.962 & 0.987  \\
		\hline
	\end{tabular}
\end{table*}

Table~\ref{tab:final_results_comparison} sets our method in context to related approaches in the literature. In general, a direct comparison of reported evaluation measures of different methods is not possible as datasets used for training and evaluation differ vastly, e.g., varying imaging quality and modality (scanned film vs. full-field digital mammography), overall number of images, or amount of training data. To counteract this issue at least to some extent, we report train and test databases in Table~\ref{tab:final_results_comparison} and refer to the respective publications for further details. In addition, we compare our results to those reported with a single (fusion) model and without augmentation at test time.

\textbf{Fusion-based methods:} 
Overall, our multi-input CNNs improve AUC scores by 0.029 to 0.04 compared to naive model ensembling. Similar increases for fusion approaches have also been reported in the literature. Kooi et al.~\cite{kooi2017} report improved AUC (+ 0.019) when adding handcrafted features (like contrast, texture) to CNN features for the classification of single mammograms. 
The work by Kyono et al.~\cite{kyono2019} fuses multi-task scores, like ``diagnosis``, ``suspicion``, ``conspicuity``, ``breast density``, across multiple views, similar to our method, but with the difference that their multi-task model predicts the same scores per view image, while we fuse predictions obtained from \textit{different} models. Adding the multi-task output to their multi-view approach increased performance by 0.031 in terms of AUC. 
Shen et al.~\cite{shengeras2021short} fuse information on a single-image level in a weakly-supervised fashion, i.e., by fusing salient image regions with a fusion module, and report a single-model AUC score of 0.833 on CBIS-DDSM test data.
A recent method by Lotter et al.~\cite{lotter2021short} combines fully and weakly (multi-instance) supervised learning and claims state-of-the-art performance for mammogram classification (AUC~=~0.963~$\pm$~0.003, OPTIMAM data). However, to obtain a score on patient level, they perform standard ensembling (average + maximum).
Finally, McKinney et al.~\cite{mckinney2020short} average cancer risk scores that are predicted by an ensemble of three large-scale deep learning models (AUC = 0.889, OPTIMAM data). Each model fuses features at different stages and aggregates predictions in various ways, e.g., by considering the maximum score or via MLPs. 

When looking at the dedicated lesion prediction, we observe that~- to the best of our knowledge - our method is the only one that specifically investigated this classification target. With an AUC score of 0.962 (F1 = 0.948), this model could be reliably used, e.g., within a reporting system, where patients with lesions are examined first. 

\textbf{Non-fusion-based methods:} Apart from the summarized information fusion methods, there are numerous works that predict whether an image is malignant directly from a view image~\cite{shen2019,shu2020}, or/and additionally apply simple ensembling strategies for predictions on breast- or patient-level~\cite{ribli2018,kim2020short}. 
Shen et al.~\cite{shen2019}, for example, utilize patch-based pre-training and compare variants of ResNet and VGG in their work. They report an image-level AUC score of 0.87 on CBIS-DDSM and 0.95 on the INbreast dataset (transfer-learned). Shu et al.~\cite{shu2020} propose two region-based pooling strategies and achieve lower AUC scores on CBIS-DDSM (0.838) and INbreast (0.934) data as compared to Shen et al.~\cite{shen2019}. 
Ribli et al.~\cite{ribli2018} localize suspicious lesions using Faster R-CNN and consider the maximum/average score on image/breast level (AUC = 0.95 on INbreast data). 

\textbf{Fusion- vs. non-fusion-based methods:} The results summarized in Table~\ref{tab:final_results_comparison} show competitive performance of fusion- and non-fusion-based methods. Although there is no clear benefit of fusion-based approaches over non-fusion-based works in terms of AUC scores, fusion approaches show different advantages. 
Recent methods focused, for example, on the integration of radiological and clinical features or aimed at increasing interpretability of models, which is an important aspect in the medical domain~\cite{kyono2019,barnett2021,shengeras2021short,lotter2021short}. These advantages, however, may come at the cost of more complex training procedures as compared to standard deep learning models~\cite{shengeras2021short}.
One limitation of recent fusion-based approaches, including this work, is the requirement for detailed, high-quality expert-annotations~\cite{kyono2020,mckinney2020short,kooi2017,barnett2021}. However, this is not limited to fusion methods per se, as the need for, e.g., bounding box annotations applies likewise to non-fusion-based, R-CNN/YOLO-based localization approaches~\cite{ribli2018,agarwal2020,akselrodballin2017,almasni2018short}. Recent weakly supervised works aim to tackle this issue and show already promising results~\cite{lotter2021short,shengeras2021short,tardymateus2021}.

\subsubsection{Clinical Implications}
\label{sec:clinical implications}

In this work, we presented a technical proof-of-concept study for a mammography pipeline comprising of three task-specific models and patient meta-models that fuse task-specific features and predictions. While one goal was to obtain an improved assessment on patient level as compared to standard model ensembling, the second goal was to develop a support tool for reading tasks of radiologists.
Similar to recent technical proof-of-concept studies by Kyono et al.~\cite{kyono2019,kyono2020} and Barnett et al.~\cite{barnett2021}, we aimed to provide intermediate results that are linked to radiological features and potential cancer risk factors. This is in contrast to studies that highlight the potential of workload reduction by excluding scans from reading that are very likely normal, i.e., do not have any suspicious lesions~\cite{kyono2020,rodriguezruiz2019_study_normals_short,ou2021,geras2019}.
However, our global lesion and malignancy predictions could be used to prioritize images for reading instead of excluding them, and additional intermediate results of task-specific models can be presented to the clinicians during exam reading and diagnosis. 
Localizing suspicious lesions, for example, is an essential part when reading mammograms where clinicians examine both views and breasts~\cite{shachor2020}. Thus, showing localized regions can aid radiologists in image interpretation~\cite{soffer2019}, for example, showing only the most important findings and raising attention for them~\cite{rodriguezruiz2019,rodriguezruiz2019_101_rad_short}. Recent user studies in other domains like prostate cancer diagnosis confirmed that prompting clinicians to suspicious regions helped them in reading~\cite{cai2019}. 
In addition to the localized regions, our pipeline estimates the patient's breast density, which is an important risk factor for developing breast cancer~\cite{destounis2020}, as already discussed in Section~\ref{sec:breast density discussion}.

\subsubsection{Limitations}

One limitation of this study is the relatively small DDSM dataset with only 2254 patients (after curation), as compared to resources used by related works~\cite{shengeras2021short,wu2018short,agarwal2020,mckinney2020short,kyono2019}. Further, as the DDSM data consists of scanned film mammograms only, the imaging quality is significantly lower as compared to full-field digital mammography images. However, the usage of a fully open dataset fosters the development and comparability of approaches, while, e.g., access to the OPTIMAM database and high-quality, expert-annotated data in general remains limited~\cite{ou2021,soffer2019}. 
A second factor may be the fixed number of detected lesions $n$ currently used in our fusion models, which could be targeted with a multi-instance approach in the future. Finally, the use of a combined model that performs lesion and malignancy prediction in a multi-task fashion would reduce the number of models and could potentially improve also the performance of malignancy prediction.

\subsubsection{Future Perspectives}

The transfer of the complete pipeline to a large, full-field digital mammography dataset would be the logical next step and potentially boost performance when trained on a larger data resource. To mitigate the requirement for expensive bounding box annotations for the lesion localization model $\LocalizationModel$, an interpretable, weak localization approach similar to our recent works~\cite{major2020,lenis2020} can be integrated in conjunction with the findings model $\FindingsModel$. Further, the improvement of the localization performance of model $\LocalizationModel$, especially for benign lesions, as well as the training of one combined model for lesion and malignancy prediction are considered future work. The inclusion of additional radiological features, such as non-image-based risk factors (e.g., patient age, patient/family history) would be of interest and importance for clinical use~\cite{weaverleung2018}. Moreover, the analysis of temporal change of lesions is considered an important biomarker in practice~\cite{sechopoulos2021,kooi2017_differences}. 
Finally, the evaluation of our proposed pipeline in a clinical reader study would help to further evaluate the potential benefits and risks of having global and local, task-specific information available, e.g., in terms of acceptance, increased interpretability, but also potential bias~\cite{ou2021}.

\section{Conclusion}
\label{sec:conclusion}

In this work, we proposed the fusion of predictions and features from different \textit{task-specific models} for improving mammography screening data classification. We trained and evaluated our fusion models for two different classification targets relevant in the field of mammogram analysis: the prediction of \textit{(i)} the presence of any lesion and \textit{(ii)} the presence of any malignant lesion in a patient. 
Our experiments on public mammography data showed that the fusion of scores with MLPs as well as feature fusion with multi-input embedding CNNs improves AUC scores compared to standard ensembling. Overall, we report an AUC score of 0.962 for predicting the presence of any lesion and 0.791 for classifying the presence of malignant lesions on patient level. By supporting our global predictions per patient with the local, sub-results obtained by the task-specific models, we aim to aid clinicians in their reading and decision process. 
Finally, we performed an ablation study with breast density scores and features and conclude that additional density information can benefit the classification performance for both target scores.

\bibliographystyle{IEEEtran}
\bibliography{arxiv_version}

\end{document}